\documentclass[aps,prd,reprint,longbibliography,floatfix,nofootinbib,balancelastpage]{revtex4-2}
\pdfoutput=1
\usepackage{amssymb,graphicx}
\usepackage{epstopdf}
\usepackage{mathrsfs}
\usepackage{centernot}
\usepackage{amssymb}
\usepackage{verbatim}
\usepackage{xcolor}
\usepackage{multirow}
\usepackage{amsmath}
\usepackage{braket}
\usepackage{subfig}
\usepackage{slashed} 
\usepackage{float}
\usepackage[normalem]{ulem}
\usepackage{sidecap}
\usepackage{hyperref}
\usepackage{cancel}
\usepackage{enumerate}
\hypersetup{pdfstartview=FitV,colorlinks=true,linkcolor=blue,citecolor=red,filecolor=black,urlcolor=blue}
\usepackage[T1]{fontenc}
\usepackage[utf8]{inputenc}
\usepackage{tikz, xcolor,hyperref}
\usepackage{amsmath, amssymb, amsthm, graphicx, epsfig, fancyhdr,epsfig, slashed}
\usepackage[normalem]{ulem}
\usepackage{natbib}
\usepackage{float}
\usepackage{tikzsymbols}
\usepackage{makecell} 

\usetikzlibrary{shapes.geometric,positioning,backgrounds}

\def\m{\mu}
\def\n{\nu}						
\def\r{\rho}	
\def\s{\sigma}

\def\be{\begin{equation}}

\def\ee{\end{equation}}
\def\bea{\begin{eqnarray}}
\def\eea{\end{eqnarray}}
\def\bi{\begin{itemize}}
\def\ei{\end{itemize}}
\def\beq{\begin{equation}\begin{aligned}}
\def\eeq{\end{aligned}\end{equation}}

\numberwithin{equation}{section}
\usepackage[compat=1.1.0]{tikz-feynman}
\usetikzlibrary{shapes.geometric}

\newcommand{\fsl}[1]{{\centernot{#1}}}
\tikzset{graviton/.style={decorate, decoration={snake, amplitude=.4mm, segment length=1.5mm, pre length=.5mm, post length=.5mm}, double}}
\tikzfeynmanset{
  anomaly/.style={
    /tikz/draw,
    /tikz/fill=gray!25,
    regular polygon,
    regular polygon sides=3,
    shape border rotate=180,
    minimum size=1.0cm,
    inner sep=1pt
  }
}

\begin{document}

\title{Gravitational production of dark QED by the trace anomaly}

\author{Simon Cléry$^{a}$}
\email{simon.clery@tum.de}
\author{Aliaksei Kachanovich$^{b}$}
\email{aliaksei.kachanovich@ulb.be}
\author{Michel HG Tytgat$^{b}$}
\email{michel.tytgat@ulb.be}

\vspace{0.1cm}

\affiliation{Technical University of Munich, TUM School of Natural Sciences, Physics Department, 85748
Garching, Germany
${}^a$
 }
 \affiliation{\,Service de Physique Th\'eorique, Universit\'e Libre de Bruxelles,
Boulevard du Triomphe, CP225, 1050 Brussels, Belgium
${}^b$ }

\begin{abstract}
We revisit the production of fermionic dark matter $\chi$ through the graviton portal, sourced by inflaton annihilation at the end of inflation. Because the production rate is helicity-suppressed, $\propto m_\chi^2$, the dark fermion must be very heavy to account for the observed relic abundance, $m_\chi \gtrsim 10^{10}$~GeV, with the exact bound set by the reheating temperature $T_{\rm RH}$. Here we examine how interactions within the dark sector could modify this result. As a toy model, we consider dark QED, in which $\chi$ couples to a light dark photon. {We first show that, as a consequence of classical conformal invariance, bremsstrahlung of a dark photon does not lift the helicity suppression for light fermions.} We then show that the trace anomaly of dark QED, which affects the coupling of the graviton to dark photons, can produce light dark fermions via a Dalitz process. This allows to extend the fermionic dark mass range possibly down to $m_\chi \sim 10^7$~GeV, depending on $T_{\rm RH}$ and the dark QED coupling. We also determine under which conditions the hidden sector reaches thermal equilibrium and derive a lower bound on the temperature ratio at reheating $(T'/T)_{\rm RH}$ as function of the dark QED coupling and the reheating temperature. 
\end{abstract}
\maketitle

\section{Introduction}
\label{sec:intro}

The nature and origin of dark matter (DM) remain unknown. A popular framework assumes that DM belongs to a hidden sector (HS), while ordinary matter resides in the so-called visible sector (VS), meaning essentially the Standard Model (SM). The two sectors may interact through a variety of portals \cite{Patt:2006fw} and can be produced through freeze-in and related mechanisms \cite{Hall:2009bx,Chu:2011be,Bernal:2017kxu}. However, irrespective of the details of these interactions, they necessarily couple through gravity. Several groups have investigated a minimal scenario in which a HS is populated only through graviton-mediated interactions \cite{Garny:2015sjg,Garny:2017kha,Bernal:2018qlk,Chung:2018ayg,Mambrini:2021zpp}. Since gravity is unavoidable, this mechanism provides in principle a lower bound on the abundance of HS particles produced in the early Universe. Because gravitational interactions are Planck-suppressed, production proceeds via UV freeze-in \cite{Elahi:2014fsa} and is therefore highly sensitive to the conditions at the earliest times. Moreover, gravitational production of DM can occur at several distinct epochs and through distinct mechanisms, from quantum fluctuations during inflation \cite{Graham2015Vector,Ema:2018ucl}, from inflaton scattering at the end of inflation \cite{Clery:2021bwz, Clery:2022wib, Barman:2022qgt} or from scattering of thermalized SM particles during reheating \cite{Bernal:2018qlk, Mambrini:2021zpp, Clery:2021bwz}. Which channel dominates depends sensitively on the DM mass and on the reheating dynamics.

In this work, we focus on the production of fermionic DM through the scattering of inflaton particles.  We consider dark QED, a simple but instructive HS model \cite{Ackerman:2008kmp,Feng:2009mn,Chu:2011be}. The DM consists of a Dirac fermion, $\chi$, coupled to a dark photon $\gamma'$, {which we assume to be massive, but lighter than the fermionic DM. We are mainly interested in the production of its transverse modes, which can source the dark fermion abundance. The longitudinal mode is also produced gravitationally, but we show that it decouples from the dark fermions and, for the parameter range we consider, remains a negligible fraction of the DM and dark radiation.} The coupling in the HS is $e'$ and $\alpha' = e'^2/4\pi$. Traditionally, dark QED is supposed to interact with the SM or visible sector (VS) through kinetic mixing \cite{Holdom:1985ag}. Here, we assume that this channel is subdominant, $\epsilon \sim 0$, and focus on the gravitational portal. As shown in \cite{Gherghetta:2019coi}, this assumption may be self-consistent for dark QED. \footnote{For an exact dark charge-conjugation symmetry $C_{\rm D}$, $F_{\mu\nu}\to -F_{\mu\nu}$, so that
the kinetic-mixing operator $B_{\mu\nu}F^{\mu\nu}$ is odd, while gravitational
couplings through the energy-momentum tensor $T_{\mu\nu}$ are even. Gravity
therefore cannot generate kinetic mixing as long as $C_{\rm D}$ is preserved. If it is broken in the UV, gravity-induced kinetic mixing can arise, but only at very high loop order. Estimates based on gravity alone require at least three-graviton exchange,
plus Higgs/dark-Higgs exchange inside the fermion loops, leading to six-loop diagrams and the estimate   $\varepsilon\sim10^{-13}$ at the Planck scale \cite{Gherghetta:2019coi}. We have verified that such a tiny mixing parameter leads to negligible production of HS particles for the DM masses and reheating temperatures considered in this work.} 

\bigskip

The setup we consider is as follows. At the end of
inflation, the inflaton ($\phi$) field oscillates about the minimum of its potential.
This coherent oscillation can be described as a condensate of highly
non-relativistic inflaton particles. During this epoch, dark fermions can be
produced gravitationally through the $s$-wave annihilation of a pair of
inflatons, $\phi \phi \to\chi\bar\chi$. This process is, however, helicity
suppressed, ${\cal M}\propto m_\chi$~\cite{Clery:2021bwz}. The reason is that only the trace of the dark-fermion stress-energy tensor
contributes in the $s$-wave, and on shell $T^\mu{}_\mu\propto m_\chi\bar\chi\chi$.\footnote{Such a suppression is not specific to the graviton portal. In the $s$-wave, a pair of spinless particles (here the inflatons) has $J=0$. If the fermions couple through a chirality-preserving interaction (vector or axial currents, or chiral Yukawa couplings to a $t$-channel mediator), the fermion and antifermion have opposite helicities in the massless limit, hence $|J_z|=1$ along their direction of motion. A $J=0$ final state then requires a chirality flip, so the $s$-wave amplitude is $\propto m_\chi$ and $\sigma v\propto m_\chi^2$. The suppression is absent when the coupling itself flips chirality, as for an $s$-channel scalar coupling to $\bar\chi\chi$. For real scalars, Bose symmetry further forbids the $p$-wave, so for $m_\chi\to0$ the leading unsuppressed contribution is $d$-wave, e.g.\ through a $t$-channel chiral coupling~\cite{Toma:2013bka,Giacchino:2013bta}.}
Combined with the Planck scale suppression of the gravitational interaction, the observed DM abundance requires a very heavy dark fermion, $m_\chi\gtrsim10^{10}\, \rm GeV$, the precise value depending on the reheating temperature $T_{\rm RH}$~\cite{Clery:2021bwz}. 

In this work, we investigate to what extent these conclusions change when the
dark fermions interact with other hidden-sector particles. In dark QED, this
role is played by the dark photon. {The gravitational production of such an interacting dark sector from the SM thermal bath, and its possible thermalization, has been studied in~\cite{Garny:2018grs}. Here we focus instead on its production from the inflaton condensate.} One could suppose that emission of a dark photon,
$\phi \phi \to\chi\bar\chi\gamma'$ could lift the helicity suppression, at the cost of
a factor $\alpha'$ and three-body phase space, in analogy with 
bremsstrahlung in the case of Majorana, and more to the point, real scalar particles annihilation~\cite{Bergstrom:1988fp,Toma:2013bka,Giacchino:2013bta}. However, as we will emphasize, this is not the case for the graviton portal because the trace of the dark QED stress-energy tensor is helicity suppressed to order ${\cal O}(e')$. We thus consider the leading unsuppressed contribution which arises from the trace anomaly, which is ${\cal O}(e'^3)$ and which
opens the channel $\phi \phi \to\gamma'\gamma'$ and, at even higher order,
$\phi \phi \to\chi\bar\chi \gamma'$ through an off-shell dark photon. We will show that, despite the extra powers of $e'$, the DM mass range can differ from the case of a HS only made of dark fermions and, in particular, could reach lower DM masses (see red curves in figures \ref{fig:relics}).

\bigskip
The article is organized as follows. After a brief introduction in section \ref{sec:framework} to the gravitational portal, we consider bremsstrahlung emission of a dark photon, $\phi\phi \rightarrow \chi\bar\chi \gamma'$ in section \ref{sec:bremsstrahlung}. Because we deem it an interesting check, we show explicitly that this process remains helicity suppressed and relate this conclusion to the fact that the stress-energy tensor is traceless at tree level for $m_\chi = 0$. In section \ref{sec:trace_anomaly}, we turn to production of dark fermions and dark photons through the trace anomaly. Provided the gauge coupling is not too small, we show that the dark fermion production may occur through the dark QED trace anomaly. We also emphasize that the suppression from extra powers of the coupling $e'$ is partly alleviated by a logarithmic enhancement in the process $\phi\phi \rightarrow \chi\bar\chi \gamma'$ when the fermion and anti-fermion are collinear, a process akin to the Dalitz decay of the neutral pion $\pi^0 \rightarrow e^+e^-\gamma$ through the chiral anomaly. We study the DM abundance produced during reheating in section \ref{sec:reheating}. There we compare several production mechanisms during reheating and show that, provided the HS gauge coupling is not too small and depending on the reheating temperature, 
the DM mass can be substantially lower, possibly down to $m_\chi \sim 10^7$ GeV. Finally, we determine under which condition the hidden sector so produced may reach thermal equilibrium at $T'$ and determine a lower bound and the ratio of temperatures at reheating, $\xi_{\rm RH} = (T'/T)_{\rm RH}$, as a function of the dark QED coupling and the reheating temperature. Feynman rules, as well as the Dalitz process kinematics and sequential production of dark fermions, are discussed in the appendices.

\section{Gravitational portal}
\label{sec:framework}
We first recap the processes of production of a pair of dark fermions ($\chi$) or a pair of dark photons ($\gamma'$) from annihilation of a pair of inflaton particles ($\phi$) through one graviton exchange ($h$), see figures \ref{Fig:fermions} and \ref{Fig:photons}. 
\begin{center}
\bea
\scalebox{1.2}{
\begin{tikzpicture}[baseline={-0.1cm}]
  \begin{feynman}[every blob={/tikz/fill=gray!30,/tikz/inner sep=2pt}]
    \vertex (i1) at (0, 0.75) {\(\phi\)} ;
    \vertex (i2) at (0,-0.75) {\(\phi\)};
    \vertex (c1) at (1., 0.) ;
    \vertex (c2) at (2., 0.) ;
    \vertex (f1) at (3.,0.75) {\(\chi\)};
    \vertex (f2) at (3.,-0.75)  {\(\bar{\chi}\)};;
    \diagram* {
      (i1) -- [scalar] (c1),
      (c1) -- [scalar] (i2),
      (c1) -- [graviton, style=black, edge label=\(h\)] (c2),
      (c2) -- [fermion] (f1),
      (f2) -- [fermion] (c2),      
      }; \end{feynman}
\end{tikzpicture}}
\label{Fig:fermions}
\eea
\end{center}

\begin{center}
\bea
\scalebox{1.2}{
\begin{tikzpicture}[baseline={-0.1cm}]
  \begin{feynman}[every blob={/tikz/fill=gray!30,/tikz/inner sep=2pt}]
    \vertex (i1) at (0, 0.75) {\(\phi\)} ;
    \vertex (i2) at (0,-0.75) {\(\phi\)};
    \vertex (c1) at (1., 0.) ;
    \vertex (c2) at (2., 0.) ;
    \vertex (f1) at (3.,0.75) {\(\gamma'\)};
    \vertex (f2) at (3.,-0.75)  {\(\gamma'\)};;
    \diagram* {
      (i1) -- [scalar] (c1),
      (c1) -- [scalar] (i2),
      (c1) -- [graviton, style=black, edge label=\(h\)] (c2),
      (c2) -- [boson] (f1),
      (f2) -- [boson] (c2),      
      }; \end{feynman}
\end{tikzpicture}}
\label{Fig:photons}
\eea
\end{center}

As usual, the graviton field $h_{\mu\nu}$ is defined as a weak-field perturbation around a flat Minkowski metric, 
$g_{\mu \nu} \simeq \eta_{\mu \nu} + 2 {h}_{\mu \nu}/M_P$ where  $h_{\mu\nu}$ is canonically normalized and $M_P  = (8\pi G)^{-1/2} \simeq 2.4 \times 10^{18} \, \rm{GeV}$ the reduced Planck mass, see e.g. \cite{Choi:1994ax,Holstein:2006bh}. To quadratic order in the graviton field and in the de Donder/harmonic gauge, the Einstein-Hilbert Lagrangian takes the form
\beq
{\cal L} = \frac{M_P^2}{2}\sqrt{-g} R \supset  \frac12 \partial^\alpha h^{\mu \nu}\partial_\alpha h_{\mu \nu} - \frac14 \partial_\mu h \partial^\mu h 
\eeq
with $h = h^\alpha_\alpha$. The propagator for a graviton with momentum $P$ is then given by 
\begin{equation}
 \Pi^{\m\n\rho\sigma}(P) = {i \over 2 P^2} \left({\eta^{\rho\m}\eta^{\sigma\n} + 
\eta^{\rho\n}\eta^{\sigma\m} - \eta^{\rho\sigma}\eta^{\m\n} }\right)
\end{equation} 
Similarly, the interaction of the graviton with both the inflaton and the HS is determined, to leading order in $h_{\mu\nu}$ {and to leading order in the coupling} ${\cal O}(e'^0)$, by 
\beq
\sqrt{-g}{\cal L} \supset -\frac{1}{M_P}h_{\mu \nu}
\left(T^{\mu \nu}_\phi+T^{\mu \nu}_\chi + T^{\mu \nu}_{\gamma'} \right) 
\label{Eq:lagrangian_free}
\eeq
where
\bea
T^{\mu \nu}_\phi &=&
\partial^\mu \phi \partial^\nu \phi-
\eta^{\mu \nu}
\left[
\frac{1}{2}\partial^\alpha \phi \partial_\alpha \phi-V(\phi)\right] 
\label{Eq:tensors}
\\
T^{\mu \nu}_\chi &=&
\frac{i}{4}
\left[
\bar \chi \gamma^\mu \overset{\leftrightarrow}{\partial^\nu} \chi
+\bar \chi \gamma^\nu \overset{\leftrightarrow}{\partial^\mu} \chi \right] \nonumber \\
&& -\eta^{\mu \nu}\left[\frac{i}{2}
\bar \chi \gamma^\alpha \overset{\leftrightarrow}{\partial_\alpha} \chi
-m_\chi \bar \chi \chi\right] \\
\label{Eq:tensorf}
T_{\gamma'}^{\mu \nu} &=& - F^\nu_{\,\,\alpha} F^{\mu \alpha} + \frac{1}{4} \eta^{\mu \nu} F^{\alpha \beta} F_{\alpha \beta} \\
&+& m_{\gamma'}^2A^\m A^\n - \frac{1}{2}\eta^{\mu \nu}m_{\gamma'}^2A_\alpha A^{\alpha} 
\label{Eq:tensorv}
\eea
Here, $V(\phi)$ is in principle the full inflaton potential. However, for annihilation of inflaton particles during reheating we only need its behavior around the minimum. Here, for the sake of the argument, we assume that it is quadratic, $V(\phi) \approx 1/2 \, m_\phi^2 \phi^2$. 
The corresponding Feynman rules are given in appendix \ref{app:FR}. 
\bigskip

We write $T_i^{\mu\nu} \rightarrow M_i^{\mu\nu}$, with $i \in\{\phi,\chi,\gamma'\}$, the matrix element in momentum space of the above stress-energy tensors. The amplitude for inflaton annihilation into a pair of dark electrons can then be expressed as
\be
i \mathcal{M}^{0\frac{1}{2}}=  M^{\frac{1}{2}}_{\m\n} \Pi^{\m\n\r\s}M^{0}_{\r\s} \equiv M^{\frac{1}{2}}_{\m\n} S^{\m\n}
\label{eq:Mdom}
\ee
with
\be
S^{\mu\nu} = {-1\over s M_P}\left(p_1^\mu p_2^\nu + p_1^\nu p_2^\mu + m_\phi^2 \eta^{\mu\nu}\right)
\ee
where $s = (p_1 + p_2)^2$. 
As the inflaton particles are non-relativistic during annihilation, $p_1 = p_2 = p = (m_\phi,\vec 0)$ and $s = 4 m_\phi^2$, such that
\be
S^{\mu\nu} \rightarrow {-1\over 4 m_\phi^2 M_P}\left(2 p^\mu p^\nu  + m_\phi^2 \eta^{\mu\nu}\right)
\label{eq:source_trace}
\ee
The first term vanishes because of conservation of the fermion stress-energy tensor, 
\be
p^\mu M_{\mu \nu}^{\frac{1}{2}}=0
\ee
and so the amplitude $\mathcal{M}^{0\frac{1}{2}}$ reduces to\cite{Mambrini:2021zpp,Clery:2021bwz}, 
\begin{eqnarray}
i \mathcal{M}^{0\frac{1}{2}}&=& - {1\over 4 M_P}\eta^{\mu\nu}  M^{\frac{1}{2}}_{\m\n} \nonumber\\
&=& i {m_\chi\over 4 M_P^2} \bar u(p_3) v(p_4)
\label{eq:soft}
\end{eqnarray}
 The helicity suppression $\propto m_\chi$ is consistent with the inflaton pair being in a $J^{PC}=0^{++}$ state. Indeed, $J=0$ forces the fermion and anti-fermion to carry equal helicities, \textit{i.e.} opposite chiralities, which requires a chirality flip.\footnote{More precisely, for a state made of a fermion and its anti-fermion with relative orbital angular momentum $L$ and total spin $S$,
$$
P=(-1)^{L+1}, \qquad C=(-1)^{L+S},
$$
where the extra $-1$ in $P$ (relative to a bosonic case) comes from the fermion and anti-fermion having opposite intrinsic parities. $J^{PC}=0^{++}$ thus corresponds to a $^3P_0$ state ($L=S=1$). Along the axis of the relative momentum (say $z$), $L_z=0$, so $J=0$ requires $S_z=\lambda_1-\lambda_2=0$. Thus, the fermion and anti-fermion carry equal helicities $\lambda_\chi = \lambda_{\bar \chi}$, which requires a chirality flip.}
Equivalently, the inflaton pair couples to the fermions through the trace of their stress-energy tensor, $T^\mu{}_\mu = m_\chi \bar\chi\chi$ on shell, which vanishes for massless fermions as a consequence of conformal invariance.

Similarly, the amplitude for the production of a pair of dark photons is \cite{Kolb:2020fwh,Garcia:2023obw},
\begin{eqnarray}
i \mathcal{M}^{01}&=& - {m_\phi^2\over s M_P}\eta^{\mu\nu}  M^{1}_{\mu\nu} \nonumber\\
&=& - i {m_{\gamma'}^2\over 2 M_P^2}\, \epsilon^\ast_1 \cdot \epsilon^\ast_2 ,
\end{eqnarray}
where $\epsilon_{1,2}$ are the polarization four-vectors of the outgoing dark photons. Again, the structure $\epsilon_1^\ast\cdot \epsilon_2^\ast$ is consistent with a $J^{PC}=0^{++}$ initial state. Here, $J=0$ requires the two dark photons to carry equal helicities, and parity selects their symmetric combination.\footnote{For two (dark) photons, $C=+1$ automatically and $P=(-1)^L$, so that $0^{++}$ corresponds to $L=S=0$ or $L=S=2$. In the helicity basis, the symmetric combination $\vert ++\rangle + \vert --\rangle$ gives the $0^{++}$ structure $\epsilon_1^\ast\cdot\epsilon_2^\ast$, while the antisymmetric one gives the $0^{-+}$ structure $(\vec\epsilon_1^{\,\ast}\times\vec\epsilon_2^{\,\ast})\cdot\hat{\vec k}$.} Again, importantly, it is proportional to $m_{\gamma'}^2$ because the stress-energy tensor of a massless gauge field is traceless.\footnote{The suppression in the massless limit only applies to transverse polarizations. Owing to the growth with energy of the longitudinal polarization vectors, $\epsilon_L \sim k/m_{\gamma'}$, the amplitude into longitudinal modes remains finite as $m_{\gamma'}\rightarrow 0$. Taking $\epsilon_1 = (k,0,0,E)/m_{\gamma'}$ and $\epsilon_2 = (k,0,0,-E)/m_{\gamma'}$ gives \cite{Kolb:2020fwh,Garcia:2023obw}
$$
i\mathcal{M}^{01}_{LL} = - i\, {2 m_\phi^2 - m_{\gamma'}^2\over 2 M_P^2} \approx - i\, {m_\phi^2\over M_P^2}.
$$
This is consistent with the equivalence theorem. At high energy, the longitudinal mode behaves as a minimally coupled scalar (the St\"uckelberg field), whose stress-energy tensor is not traceless. Consequently, the limit $m_{\gamma'}\rightarrow 0$ of a massive dark photon does not coincide with a strictly massless one, which has only two transverse polarizations. In the sequel, unless explicitly stated, by massless dark photon we mean $m_{\gamma'}=0$.}

A consequence of the dependence of the above amplitudes on the mass of the final-state particles is that the production of dark QED particles through graviton exchange is significant only for heavy states \cite{Clery:2021bwz}. To account for the observed DM abundance, the mass of the dark fermions must then be of order $m_\chi \gtrsim 10^{10}$ GeV, the precise value depending on the reheating temperature $T_{\rm RH}$ (see section~\ref{subsec:DM_abundance} and figure~\ref{fig:relics}). In the sequel, we study to what extent the required DM mass is modified when the coupling of the dark fermions to a dark photon is taken into account.

\section{Dark photon Bremsstrahlung}
\label{sec:bremsstrahlung}

The helicity suppression of the $\mathcal{M}^{0\frac{1}{2}}$ amplitude suggests considering the bremsstrahlung process $\phi\phi \rightarrow \chi \bar\chi \gamma'$. Indeed, radiative corrections are a standard way to lift helicity suppressions, at the price of extra phase-space and gauge coupling factors, as for instance in the s-wave annihilation of Majorana DM into light fermions (see e.g.\ \cite{Bergstrom:1988fp,Rudaz:1989ij,Toma:2013bka,Giacchino:2013bta} for works from the perspective of DM abundance and signatures). However, since for scalars at rest the amplitude only probes the trace of the stress-energy tensor of the final state, the bremsstrahlung amplitude also vanishes in the limit $m_\chi, m_{\gamma'} \rightarrow 0$. Indeed, at tree level, the stress-energy tensor of dark QED is traceless in the massless limit. For the sake of completeness and because we find it an instructive exercise, we verify that the stress-energy tensor satisfies the Ward identity $\langle \chi\bar\chi \gamma'\vert T^\mu{}_\mu\vert 0\rangle = 0$  at tree level.

The amplitudes we want to evaluate are the following:
\begin{center}
\scalebox{0.9}{
\begin{tikzpicture}[baseline={-0.1cm}]
  \begin{feynman}[every blob={/tikz/fill=gray!30,/tikz/inner sep=2pt}]
    \vertex (i1) at (0, 0.75) {\(\phi\)} ;
    \vertex (i2) at (0,-0.75) {\(\phi\)};
    \vertex (c1) at (1., 0.) ;
    \vertex (c2) at (2, 0.) ;
    \vertex (c3) at (2.75, 0.75) ;
    \vertex (f1) at (3.5, 1.5) {\(\chi\)};
    \vertex (f2) at (3,-1)  {\(\bar{\chi}\)};
    \vertex (f3) at (3.5, 0.) {\(\gamma'\)};
    \diagram* {
      (i1) -- [scalar] (c1),
      (c1) -- [scalar] (i2),
      (c1) -- [graviton, style=black, edge label=\(h\)] (c2),
      (c2) -- [fermion] (c3),
      (c3) -- [fermion] (f1),
      (f2) -- [fermion] (c2),
      (f3) -- [boson] (c3),
      }; \end{feynman}
\end{tikzpicture}}\, + \,
\scalebox{0.9}{
\begin{tikzpicture}[baseline={-0.1cm}]
  \begin{feynman}[every blob={/tikz/fill=gray!30,/tikz/inner sep=2pt}]
    \vertex (i1) at (0, 0.75) {\(\phi\)} ;
    \vertex (i2) at (0,-0.75) {\(\phi\)};
    \vertex (c1) at (1., 0.) ;
    \vertex (c2) at (2, 0.) ;
    \vertex (c3) at (2.75, -0.75) ;
    \vertex (f1) at (3, 1) {\(\chi\)};
    \vertex (f2) at (3.5,-1.5)  {\(\bar{\chi}\)};
    \vertex (f3) at (3.5, 0.) {\(\gamma'\)};
    \diagram* {
      (i1) -- [scalar] (c1),
      (c1) -- [scalar] (i2),
      (c1) -- [graviton, style=black, edge label=\(h\)] (c2),
      (c2) -- [fermion] (f1),
      (c3) -- [fermion] (c2),
      (f2) -- [fermion] (c3),
      (f3) -- [boson] (c3),
      }; \end{feynman}
\end{tikzpicture}} \\
+ \scalebox{0.9}{
\begin{tikzpicture}[baseline={-0.1cm}]
  \begin{feynman}[every blob={/tikz/fill=gray!30,/tikz/inner sep=2pt}]
    \vertex (i1) at (0, 0.75) {\(\phi\)} ;
    \vertex (i2) at (0,-0.75) {\(\phi\)};
    \vertex (c1) at (1., 0.) ;
    \vertex (c2) at (2, 0.) ;
    \vertex (f1) at (3., 1.) {\(\chi\)};
    \vertex (f2) at (3.,-1.)  {\(\bar{\chi}\)};
    \vertex (f3) at (3.5, 0.) {\(\gamma'\)};
    \diagram* {
      (i1) -- [scalar] (c1),
      (c1) -- [scalar] (i2),
      (c1) -- [graviton, style=black, edge label=\(h\)] (c2),
      (c2) -- [fermion] (f1),
      (f2) -- [fermion] (c2),
      (f3) -- [boson] (c2),
      }; \end{feynman}
\end{tikzpicture}} 
+\,
\scalebox{0.9}{
\begin{tikzpicture}[baseline={-0.1cm}]
  \begin{feynman}[every blob={/tikz/fill=gray!30,/tikz/inner sep=2pt}]
    \vertex (i1) at (0, 0.75) {\(\phi\)} ;
    \vertex (i2) at (0,-0.75) {\(\phi\)};
    \vertex (c1) at (1., 0.) ;
    \vertex (c2) at (2., 0.) ;
    \vertex (c3) at (2.75,0.75); 
    \vertex (c4) at (3, -1) {\(\gamma'\)};
    \vertex (f1) at (3.5, 1.5) {\(\chi\)};
    \vertex (f2) at (3.5,0)  {\(\bar{\chi}\)};;
    \diagram* {
      (i1) -- [scalar] (c1),
      (c1) -- [scalar] (i2),
      (c1) -- [graviton, style=black, edge label=\(h\)] (c2),
      (c2) -- [boson] (c3),
      (c3) -- [fermion] (f1),
      (f2) -- [fermion] (c3),
      (c2) -- [boson] (c4),      
      }; \end{feynman}
\end{tikzpicture}}
\, 
\label{Fig:bremsstrahlung}
\end{center}
The first two and fourth amplitudes are induced by the coupling of the free fermion and photon stress-energy tensor to the graviton. The third one arises from the covariant derivative contribution to the dark QED stress-energy tensor, which leads to the following interaction terms  to $\mathcal{O}(e'^1)$
\bea
\sqrt{-g}\mathcal{L}_{\rm int} &\supset& 
 -\frac{e'}{2 M_P}h_{\m\n}\left(A^{\m}\bar{\chi}\gamma^\n\chi + A^{\n}\bar{\chi}\gamma^\m\chi\right.\nonumber\\ && \left.-2 \eta^{\m\n}A_\alpha\bar{\chi}\gamma^\alpha\chi \right) 
\label{Eq:lagrangian_coupled}
\eea

In what follows, we keep the fermion mass explicitly but set $m_{\gamma'}=0$. The first two amplitudes take the form
\bea
i \mathcal{M}_1 &=& \left(\frac{1}{M_P}\right) (ie') \; S^{\m\n} \times\label{eq:M1}\\
\nonumber
&&\bar{u}(p_3)\gamma^\alpha\frac{\fsl{l_3}+m_\chi}{l_3^2-m_\chi^2} M^{1/2}_{\m\n,1} v(p_4)~\epsilon^*_\alpha(k)\, ,\\
i \mathcal{M}_2 &=& \left(\frac{1}{M_P}\right) (ie')\;S^{\m\n}\times \label{eq:M2}\\
\nonumber
&&\bar{u}(p_3) M^{1/2}_{\m\n,2}\frac{\fsl{l_4}+m_\chi}{l_4^2-m_\chi^2}\gamma^\alpha~v(p_4)~\epsilon^*_\alpha(k)\, 
\eea
where $p_3, p_4$ are respectively the momenta of outgoing on-shell fermions and anti-fermions, $k$ the momentum carried away by the emitted dark photon, while $l_3= p_3 +k$ and $l_4=-p_4 -k$. We also need the following fermion partial amplitude
\be
M^{1/2}_{\m\n} =  \frac{1}{4} \left( \gamma_\mu K_{\nu} + \gamma_\nu K_{\mu} \right) - \eta_{\m\n}\left(\frac{1}{2}\fsl{K} -m_\chi\right)
\ee
with $K\equiv K^{(1)}= p_3-p_4+k$ in $\mathcal{M}_1$ or $K\equiv K^{(2)}= p_3-p_4-k$ in $\mathcal{M}_2$.

From the expression of the inflaton source term in the s-wave limit, Eq.\eqref{eq:source_trace}, we may check that the sum of all 4 amplitudes satisfies the Ward identity corresponding to the conservation of the dark QED stress-energy tensor to $\mathcal{O}(e')$. Hence, the only surviving part comes from its trace. With
\be
S^{\m\n}\rightarrow -{\eta^{\m\n}\over 4 M_P}
\label{eq:sourcetraceonly}
\ee
the fourth amplitude is traceless for $m_{\gamma'} = 0$. Indeed, 
\bea
\mathcal{M}_4 &=&\left( \frac{-i}{M_P}\right) e'~S^{\m\n}\\ 
\nonumber
&&\bar{u}_3\left\{ \frac{\fsl\epsilon^\ast}{2q^2}\left(k_\n q_\m+k_\m q_\n-\eta_{\m\n}k\cdot q \right) \right. \\
\nonumber 
&& \left. -\frac{\fsl k^\ast}{2q^2}\left(\epsilon^\ast_\n q_\m+\epsilon^\ast_\m q_\n-\eta_{\m\n} \epsilon^\ast\cdot q \right)\right.\\
\nonumber 
&& \left. +\frac{k\cdot q}{2q^2}\left( \gamma_\n \epsilon^\ast_\m + \gamma_\m \epsilon^\ast_\n \right) \right. \\
&& \left. -\frac{\epsilon^\ast \cdot q}{2q^2}\left( \gamma_\n k_\m + \gamma_\m k_\n \right) \right\}v_4 \equiv 0
\eea
where $q\equiv p_3+p_4$. 

Taking the trace of  the first two amplitudes gives
\be 
i \mathcal{M}_1 =  -i {e'\over 4 M_P^2}\bar u_3\left( m_\chi{\epsilon^\ast\cdot p_3\over p_3\cdot k}  - {3\over 2} \fsl{\epsilon}^\ast + \frac{1}{2} m_\chi {\fsl{\epsilon}^\ast \fsl{k}\over p_3\cdot k}\right) v_4 
\ee
and
\be 
i \mathcal{M}_2 = -i {e'\over 4 M_P^2}\bar u_3\left(- m_\chi{\epsilon^\ast\cdot p_4\over p_4\cdot k}  - {3\over 2} \fsl{\epsilon}^\ast+ \frac{1}{2}m_\chi {\fsl{\epsilon}^\ast \fsl{k}\over p_4\cdot k}\right) v_4
\ee
Finally, the third amplitude reads
\bea
i \mathcal{M}_3 &=& - \left(\frac{i e^\prime}{M_P}\right)~S^{\m\n} \bar{u}_3\left( \frac{1}{2}\gamma_\mu\epsilon^\ast_\nu+\frac{1}{2}\gamma_\nu\epsilon^\ast_\mu-\fsl{\epsilon}^\ast\eta_{\mu\nu}\right) v_4\nonumber\\
&=& -3 i \frac{ e^\prime}{4 M_P^2}\,\bar u_3 \fsl{\epsilon}^\ast v_4
\eea
This amplitude cancels the second term of the first two amplitudes, making their sum gauge invariant. 

Together, the three amplitudes include the amplitude for bremsstrahlung of a soft dark photon, $ k \rightarrow 0$, which takes the standard factorized form \cite{Peskin:1995ev}
\be
i \mathcal{M}_{\rm soft} =  -i e^\prime \left({p_3\cdot \epsilon^\ast\over p_3 \cdot k}- {p_4\cdot \epsilon^\ast\over p_4 \cdot k}\right) \mathcal{M}^{0\frac{1}{2}}
\ee
with $\mathcal{M}^{0\frac{1}{2}} \propto m_\chi$ given by Eq.\eqref{eq:soft}. The remaining terms are gauge invariant and relevant for emission of a hard dark photon, $\mathcal{O}(m_\chi k)$. {For $m_{\gamma'}\ll m_\chi$, the dark photon mass only induces corrections of order $m_{\gamma'}^2/m_\chi^2$ to these amplitudes.}
Altogether, bremsstrahlung of massless dark photons is not enough to lift the helicity suppression of Eq.\eqref{eq:soft}. 

\section{Gravitational portal with the trace anomaly}
\label{sec:trace_anomaly}

The discussion of the previous section suggests considering the production of light DM fermions through the trace anomaly. The production of massless particles from a
conformally flat gravitational background is an old topic ~\cite{Dolgov:1981nw,Kolb:2020fwh, Kolb:2023ydq}. Recent references include gauge boson production during reheating through the trace anomaly in the case of non-minimal inflaton-curvature coupling~\cite{Watanabe:2010vy} or from the decays of secluded scalars into the SM \cite{Nakayama:2025xkn}. Closer to our framework, gravitational production of a conformal non-abelian dark sector has been studied in \cite{Redi:2020ffc}.
 Here, we specifically consider the impact of the trace anomaly on the production  through the oscillating inflaton condensate of dark photons. Furthermore, we extend it to the production of dark fermions through off-shell dark photons, or Dalitz process, and study their impact on the DM mass and reheating temperature ranges.

The trace anomaly can be derived from variation of a non-local action \cite{Giannotti:2008cv}. However, this is unnecessary for our purpose as, given Eq.\eqref{eq:source_trace} or \eqref{eq:sourcetraceonly}, we only need the actual expression of the trace anomaly which,  in the massless limit $m_\chi,m_{\gamma'} \rightarrow 0$, reads ~\cite{Adler:1976zt,Collins:1976yq}
\begin{equation}
    T^\mu_{\ \mu} = \frac{\beta(e')}{2e'} F_{\rho\sigma}F^{\rho\sigma}
    \label{eq:trace_anomaly}
\end{equation}
where, for $N_\chi$ copies of Dirac fermions of unit charge under an abelian gauge group, the $\beta$ function is 
\begin{equation}
\label{eq:trace_ano}
    \beta(e')= \frac{N_\chi e'^3}{12\pi^2}.
\end{equation}
at one-loop. 

The anomalous  $\langle TJJ\rangle$ vertex can be expressed as
\begin{eqnarray}
\label{eq:the_pole}
 V^{\mu\nu\alpha\beta}_{\rm anom}(P, k_1, k_2)&=&  \frac{i}{M_P} \frac{2\beta(e')}{3e'} \frac{P^2\eta^{\mu\nu}-P^\mu P^\nu}{P^2}\nonumber \\
    && \times \left[k_1\cdot k_2\,\eta^{\alpha\beta} - k_2^\alpha k_1^\beta \right]
\end{eqnarray}
with $P = k_1 + k_2$. Taking the trace of this expression, the dilaton-like pole at  $P^2=0$ (or anomaly pole) precisely reproduces the content of \eqref{eq:trace_ano} \cite{Dolgov:1971ri,Horejsi:1997yn,Giannotti:2008cv} (see also \cite{Huang:1982ik}). 
The full one-loop expression of the vertex $\langle TJJ\rangle$ contains additional transverse-traceless form factors \cite{Giannotti:2008cv, Armillis:2009pq}. However, because of \eqref{eq:source_trace},  these form factors do not contribute to the production of massless dark photons. More generally, the leading contribution to the trace anomaly in the conformal limit is due to the anomaly pole \cite{Giannotti:2008cv, Armillis:2009pq}. Corrections not fixed by the anomaly are associated with explicit fermion-mass terms in the trace, and are suppressed for $P^2 \gg m^2_\chi$. As $P^2 = 4m_\phi^2 \gg m_\chi^2$,  we retain only the anomalous contribution.\footnote{More precisely, dispersive analyses show that for massive fermions there is no pole at $P^2=0$~\cite{Dolgov:1971ri,Horejsi:1997yn,Giannotti:2008cv}. Instead, the imaginary part of the amplitude, $I(P^2;m_\chi)$, is a smooth function with support at $P^2\geq 4m_\chi^2$ and a width $\sim m_\chi^2$. Its integral, however, is independent of $m_\chi$ and gives the anomaly. In the limit $m_\chi\to0$, $I(P^2;m_\chi)\propto\delta(P^2)$, which corresponds to a $1/P^2$ pole, i.e.\ a massless dilaton-like intermediate state. Such a state exists only if the theory has truly massless states in the IR, which underlies anomaly-matching arguments~\cite{tHooft:1979rat,Schwimmer:2010za}. For $P^2\ll m_\chi^2$, the anomaly remains an exact operator relation, but the explicit mass term cancels it, $m_\chi\bar\chi\chi\to-(\beta(e')/2e')F_{\mu\nu}F^{\mu\nu}$, so the heavy fermion decouples from $T^\mu_{\ \mu}$. This is the origin of the low-energy theorems for the Higgs couplings to photons and gluons, where the Higgs couples to $m\bar\psi\psi$~\cite{Shifman:1979eb}. For $P^2\gg m_\chi^2$, as considered here, the amplitude behaves as in the massless limit, up to $\mathcal{O}(m_\chi^2/P^2)$ corrections.}

\subsection{Anomalous production of dark photons}
\label{subsec:anomaly_dark_photons}

The most direct consequence of the trace anomaly is the production of a pair of light dark photons through the gravitational portal,
\begin{equation}
    \phi(p_1)\phi(p_2)
    \to h^*
    \to \gamma'(k_1)\gamma'(k_2)
\end{equation}

\begin{center}
\scalebox{1.0}{
\begin{tikzpicture}[baseline={-0.1cm}]
  \begin{feynman}

    \vertex (i1) at (0, 0.75) {\(\phi\)};
    \vertex (i2) at (0,-0.75) {\(\phi\)};
    \vertex (c1) at (1,0);

    \vertex (anom) at (2.4,0);

    \vertex (g1) at (4.0, 0.8) {\(\gamma'\)};
    \vertex (g2) at (4.0,-0.8) {\(\gamma'\)};

    \diagram*{
      (i1) -- [scalar] (c1),
      (i2) -- [scalar] (c1),

      (c1) -- [graviton, edge label=\(h\)] (anom),

      (anom) -- [boson] (g1),
      (anom) -- [boson] (g2),
    };

    \node[
      circle,
      draw,
      fill=gray!30,
      minimum size=5mm,
      inner sep=0pt
    ] at (anom) {\(\mathcal A\)};

  \end{feynman}
\end{tikzpicture}
}
\end{center}
with amplitude 
\begin{equation}
\mathcal{M}^{0 1}_{\rm TA} = - i {\beta(e')\over e'} {1\over 2  M_P^2}\left(k_1\cdot k_2 \,\epsilon_1^\ast\cdot \epsilon_2^\ast - k_1\cdot \epsilon_2^\ast \,k_2 \cdot \epsilon_1^\ast\right)
\end{equation}
where the subscript refers to the trace anomaly (TA). 
For transverse modes\footnote{For longitudinal modes, $k_1\cdot\epsilon_2^\ast = k_2\cdot\epsilon_1^\ast\neq0$, but the anomalous amplitude is suppressed by $m_{\gamma'}^2/m_\phi^2$.} produced from non-relativistic inflaton annihilation,  $k_1\cdot \epsilon_2^\ast = k_2 \cdot \epsilon_1^\ast = 0$ and $k_1\cdot k_2 = 2 m_\phi^2$,
\begin{equation}
\mathcal{M}^{0 1}_{\rm TA} = - i {\beta(e')\over e'} {m_\phi^2\over M_P^2} \,\epsilon_1^\ast\cdot \epsilon_2^\ast
\label{eq:TA_photons_amplitude}
\end{equation}
Summing over the two physical polarizations, we obtain
\begin{equation}
    \sum_{\rm pol.}
    \left|
        \mathcal{M}_{\rm TA}^{01}
    \right|^2
    =
    \frac{N_\chi^2\alpha'^2m_\phi^4}
         {9\pi^2M_P^4}
    \label{eq:anom_AA_squared}
\end{equation}

It will be useful in the following to introduce the two-body
phase-space integral
\begin{equation}
    I_{\gamma'}
    \equiv
    \frac{1}{2!}
    \int d\Phi_2\,
    \sum_{\rm pol.}
    \left|
        \mathcal{M}_{\rm TA}^{01}
    \right|^2 ,
    \label{eq:IAA_definition}
\end{equation}
where the factor $1/2!$ accounts for the two identical dark photons
in the final state. For two massless particles,
\begin{equation}
    \int d\Phi_2 = \frac{1}{8\pi},
\end{equation}
and therefore
\begin{equation}
    I_{\gamma'}
    =
    \frac{N_\chi^2\alpha'^2m_\phi^4}
         {72\pi^3M_P^4}
    \label{eq:IAA}
\end{equation}
The anomalous gravitational portal thus provides an irreducible source of dark photon radiation even in the limit of a massless dark photon.

\subsection{Anomalous dark fermion production}
\label{subsec:anomaly_dalitz}

One of the two gauge bosons may be off-shell and convert into a dark fermion-antifermion pair,
\begin{equation}
    \phi(p_1)\phi(p_2)
    \to h^*
    \to
    \gamma'(k)\gamma'^*(q)
    \to
    \gamma'(k)\chi(p_3)\bar{\chi}(p_4),
    \label{eq:dalitz_process}
\end{equation}
where
\begin{equation}
    q=p_3+p_4,
    \quad
    P=p_1+p_2=q+k,
    \quad
    P^2=s=4m_\phi^2
\end{equation}

\begin{center}
\scalebox{1.0}{
\begin{tikzpicture}[baseline={-0.1cm}]
  \begin{feynman}

    \vertex (i1) at (0, 0.75) {\(\phi\)};
    \vertex (i2) at (0,-0.75) {\(\phi\)};
    \vertex (c1) at (1,0);

    \vertex (anom) at (2.4,0);

    \vertex (g) at (4.0,0.9) {\(\gamma'\)};
    \vertex (c2) at (3.8,-0.7);

    \vertex (f1) at (5.0,-0.2) {\(\chi\)};
    \vertex (f2) at (5.0,-1.2) {\(\bar\chi\)};

    \diagram*{
  (i1) -- [scalar] (c1),
  (i2) -- [scalar] (c1),

  (c1) -- [graviton, edge label=\(h\)] (anom),

  (anom) -- [boson] (g),
  (anom) -- [boson] (c2),

  (c2) -- [fermion] (f1),
  (f2) -- [fermion] (c2),
};

\node at (3.1,-0.75) {\(\gamma'^*\)};

\node[
  circle,
  draw,
  fill=gray!30,
  minimum size=5mm,
  inner sep=0pt
] at (anom) {\(\mathcal A\)};

  \end{feynman}
\end{tikzpicture}
}
\end{center}
Contrary to the tree-level bremsstrahlung process discussed in section~\ref{sec:bremsstrahlung}, this process is induced by the anomalous $h\gamma'\gamma'$ vertex and is therefore not helicity suppressed.
It is also analogous to the Dalitz decay of a neutral pion, $\pi_0 \rightarrow \gamma e^+e^-$ \cite{Dalitz:1951aj,Kroll:1955zu}, which, in this case, can be related to the chiral anomaly, see \cite{Landsberg:1985gaz}. \\

Contracting the dark photon propagator and the Dark QED current
with the anomaly-induced vertex gives the following amplitude for the Dalitz process,
\begin{equation}
    i\mathcal{M}^{0\frac{1}{2}}_{\rm D}
    =
    i\,\frac{\beta(e')}{2M_P^2}\,
    \frac{1}{q^2}
    \left(
        q\cdot k\,\eta^{\alpha\beta}
        -
        q^\beta k^\alpha
    \right)
    \epsilon_\beta^*(k)\,
    \bar{u}(p_3)\gamma_\alpha v(p_4)
    \label{eq:dalitz_amplitude}
\end{equation}
The explicit factor of $e'$ associated with the
$\gamma'^*\chi\bar{\chi}$ vertex cancels the $1/e'$ entering the
anomalous $\gamma'\gamma'$ vertex, such that the amplitude scales as $\beta(e')\propto e'^3$. Moreover, Eq.~\eqref{eq:dalitz_amplitude} is transverse with respect to both gauge-boson momenta, as required by Ward identities from gauge invariance.\\

We then perform the phase-space integration over the three-body final state,
\begin{equation}
    I_\chi^{\rm D}
    \equiv N_\chi
    \int d\Phi_3\,
    \sum_{\rm spin,\, pol.}|\mathcal{M}^{0\frac{1}{2}}_{\rm D}|^2 
\end{equation}
{where we consider $N_\chi$ copies of degenerate Dirac fermions in the final state.} To this purpose, we introduce the following kinematic variables,
\begin{equation}
    r\equiv\frac{m_\chi}{m_\phi},
    \qquad
    z\equiv\frac{q^2}{s},
\end{equation}
with $r^2\leq z\leq1$. The three-body phase space integral and the corresponding kinematic boundaries are derived in
Appendix.~\ref{appendix_3body}. In particular,
Eq.~\eqref{eq:dID_dz_appendix} gives
\begin{equation}
    \frac{d I_\chi^{\rm D}}{dz}
    =
    N_\chi \frac{m_\phi}{32\pi^3}
    \int_{E_3^-}^{E_3^+}
    dE_3\,
    \sum_{\rm spin,\, pol.} |\mathcal{M}^{0\frac{1}{2}}_{\rm D}|^2 ,
    \label{eq:dI3_dz_start}
\end{equation}
where the integration boundaries $E_3^\pm(z)$ are given in
Eq.~\eqref{eq:E3_boundaries}. 
After summing over final-state spins and polarizations and over the final state fermions angular variable, we obtain,
\begin{equation}
    \frac{d I_\chi^{\rm D}}{dz}
    =
    \frac{N_\chi^3 \alpha'^3m_\phi^4}
         {108\pi^4M_P^4}
    \frac{(1-z)^3}{z}
    \left(
        1+\frac{r^2}{2z}
    \right)
    \sqrt{1-\frac{r^2}{z}},
    \label{eq:Dalitz_integral}
\end{equation}

The structure of Eq.~\eqref{eq:Dalitz_integral} admits a simple
interpretation: using the two-body dark-photon production result obtained in Eq.~\eqref{eq:IAA}, the three-body Dalitz integral can be written as
\begin{equation}
    \frac{d I_\chi^{\rm D}}{dz}
    =
    I_{\gamma'}
    \frac{2 N_\chi \alpha'}{3\pi}
    \frac{(1-z)^3}{z}
    \left(
        1+\frac{r^2}{2z}
    \right)
    \sqrt{1-\frac{r^2}{z}}\,
    \label{eq:dalitz_factorized}
\end{equation}
This highlights the standard structure of an internal conversion of one of the anomaly-produced gauge bosons into a fermion-antifermion pair \cite{Landsberg:1985gaz, Kroll:1955zu}.
The additional factor $\alpha'$ associated with the conversion explains
the scaling $I_\chi^{\rm D}\propto \alpha'^3$, compared with
$I_{\gamma'}\propto \alpha'^2$.

Finally, integrating Eq.~\eqref{eq:Dalitz_integral} over the remaining kinematic variable give, 
\begin{equation}
    I_\chi^{\rm D}
    =
    \frac{N_\chi^3 \alpha'^3m_\phi^4}
         {144\pi^4M_P^4}
    J(r),
    \label{eq:I3_result}
\end{equation}
where
\begin{equation}
    J(r)
    =
    \frac{2}{3}
    \int_{r^2}^{1}
    dz\,
    (1-z)^3
    (r^2+2z)
    \frac{\sqrt{z-r^2}}{z^{5/2}} 
    \label{eq:J_definition}
\end{equation}
The integral can be performed analytically, yielding
\begin{align}
    J(r)
    ={}&
    \frac{\sqrt{1-r^2}}{3}
    \left(r^4+13r^2-14\right)
    \nonumber\\
    &+
    \left(
        \frac{r^6}{6}
        -\frac{3r^4}{2}
        +\frac{4}{3}
    \right)
    \log\left(
        \frac{1+\sqrt{1-r^2}}
             {1-\sqrt{1-r^2}}
    \right)
    \label{eq:J_exact}
\end{align}
For light fermions, i.e $r\equiv m_\chi/ m_\phi \ll1$, it reduces to
\begin{equation}
    J(r)
    =
    \frac{8}{3}\log\frac{2}{r}
    -\frac{14}{3}
    +\mathcal{O}(r^2)
    =
    \frac{4}{3}
    \left[
        \log\frac{4}{r^2}
        -\frac{7}{2}
    \right]
    +\mathcal{O}(r^2)
    \label{eq:J_smallr}
\end{equation}
The logarithmic behavior is due to the collinear enhancement associated with the conversion of an almost on-shell virtual gauge boson into a light fermion-antifermion pair \cite{Kroll:1955zu}. The singular limit $q^2\to0$ is regulated by the physical threshold $q^2\geq4m_\chi^2$.\\

The anomaly-induced Dalitz process is 1-loop suppressed but, in contrast with both the minimal two-body channel and the tree-level bremsstrahlung contribution, it is not helicity suppressed by the factor $m_\chi^2/m_\phi^2$. This makes it potentially relevant for light dark fermions, $m_\chi/m_\phi \ll 1$, produced in the early Universe.

\section{Cosmological abundances}
\label{sec:reheating}

\subsection{Production rates during reheating}
\label{subsec:production_rates}

We now turn to the cosmological production of dark sector particles
during the phase of reheating after the end of inflation. As stated before, we assume that, close to the minimum of the inflaton potential, $V(\phi)\simeq \frac{1}{2}m_\phi^2\phi^2$, such that the oscillating inflaton condensate behaves as a matter-like, pressure-less component. During this stage, the condensate is spatially homogeneous. Its zero mode may be described by a phase-space distribution sharply peaked at vanishing spatial momentum,
\begin{equation}
    f_\phi(\mathbf{k},t)
    =
    (2\pi)^3 n_\phi(t)\,
    \delta^{(3)}(\mathbf{k}),
\end{equation}
with
\begin{equation}
    \rho_\phi=m_\phi n_\phi
\end{equation}
The gravitational portals processes considered in the previous sections are
then sourced by the second harmonic of the oscillating condensate,
corresponding to a center-of-mass energy
\begin{equation}
    \sqrt{s}=2m_\phi
\end{equation}
The results derived in section~\ref{sec:trace_anomaly} can be translated into source terms from the oscillating inflaton condensate. Following the classical-condensate treatment of \cite{Garcia:2020wiy}, the relevant production rates of species $X$ from the quadratic inflaton condensate can then be written schematically as~\footnote{For the coherent quadratic inflaton condensate, the second harmonic gives an event rate $R_{\rm event} =\frac14 n_\phi^2\sigma v_{\rm rel}$ compared to the standard annihilation cross section, as derived in Appendix~\ref{appendix_3body}.
This differs by a factor of two from the dilute-gas expression
$\frac12 n_\phi^2\sigma v_{\rm rel}$ for identical initial particles.} 
\begin{equation}
    R
    =
    \mathcal N\frac{\rho_\phi^2}{16m_\phi^4}\,
    \int d\Phi\,|\mathcal M|^2 ,
    \label{eq:condensate_rate_general}
\end{equation}
{with $\mathcal N$ the number of particles of one species produced in the process. We consider different species to be produced and use the labels $\chi$ (counting both $\chi$ and $\bar\chi$), $\gamma'_T$ or $\gamma'_L$ for the transverse and longitudinal dark photon modes. When several channels produce the same species, a superscript specifies the channel: "${\rm D}$" for the anomaly-induced Dalitz process, "${\rm min}$" for the minimal gravitational process, "${\rm bath}$" for gravitational production from the SM thermal bath, and "${\rm seq}$" for sequential production $\gamma'\gamma'\to\chi\bar\chi$.}

For the anomalous gravitational portal producing two transverse dark photons, we then have
\begin{equation}
    R_{\gamma'_T}
    =
    \frac{\rho_\phi^2}{8m_\phi^4}
    I_{\gamma'}
    \label{eq:Rgamma_from_I}
\end{equation}
as $\mathcal N_{\gamma'}=2$. Using Eq.~\eqref{eq:IAA}, we obtain \cite{Dolgov:1981nw}
\begin{equation}
    R_{\gamma'_T}
    =
    \frac{N_\chi^2 \alpha'^2\rho_\phi^2}
         {576\pi^3M_P^4}
    \label{eq:Rgamma}
\end{equation}
Similarly, each Dalitz event produces one fermion-antifermion pair, with $\mathcal N_\chi =2$, the sum over the $N_\chi$ copies of Dirac fermions being included in $I_\chi^{\rm D}$. Defining $n_\chi$ in the following as the total
number density of $\chi+\bar\chi$, the corresponding source is
\begin{equation}
    R_\chi^{\rm D}
    =
    \frac{\rho_\phi^2}{8m_\phi^4}
    I_\chi^{\rm D} 
\end{equation}
Using Eq.~\eqref{eq:Dalitz_integral}, we find
\begin{equation}
    R_\chi^{\rm D}
    =
    \frac{N_\chi^3 \alpha'^3\rho_\phi^2}
         {1152\pi^4M_P^4}
    J(r)
    \label{eq:Rdalitz}
\end{equation}
which, for light fermions, $r= m_\chi/m_\phi\ll1$, takes the form
\begin{equation}
    R_\chi^{\rm D}
    \simeq
    \frac{N_\chi^3 \alpha'^3\rho_\phi^2}
         {864\pi^4M_P^4}
    \left[
        \log\frac{4m_\phi^2}{m_\chi^2}
        -\frac{7}{2}
    \right]
    \label{eq:Rdalitz_smallr}
\end{equation}
{For comparison, the minimal gravitational portal production of a Dirac
fermion from the inflaton condensate is given by \cite{Clery:2021bwz}
\begin{equation}
    R_\chi^{\rm min}
    =
    \frac{N_\chi\rho_\phi^2}{128\pi M_P^4}
    \left({m_\chi\over m_\phi}\right)^2\left(1-\left({m_\chi\over m_\phi}\right)^2\right)^{3/2}
    \label{eq:Rminimal}
\end{equation}}
The different behavior in the light fermion limit is manifest. While the minimal gravitational source is helicity suppressed, $R_\chi^{\rm min}\propto (m_\chi/m_\phi)^2$, the anomaly-induced contribution remains finite up to the logarithmic enhancement associated with the Dalitz conversion, $R_\chi^{\rm D} \propto \alpha'^3\log\frac{1}{r^2}$. By computing the ratio of the two rates in the limit $r\ll 1$, we find,
\begin{equation}
\frac{R_\chi^{\rm D}}{R_\chi^{\rm min}}
\simeq
\frac{4N_\chi^2\alpha'^3}{27\pi^3}
\frac{
\log(4/r^2)-7/2
}{r^2}
\end{equation}

Consequently, despite its loop suppression, the anomalous channel
can dominate over the minimal gravitational portal for sufficiently small fermion masses and/or sufficiently large dark sector coupling $e'$, more precisely under the condition,
\begin{equation}
\frac{m_\chi^2}{m_\phi^2}
\lesssim
\frac{4N_\chi^2\alpha'^3}{27\pi^3}
\left[
\log\frac4{r^2}-\frac72
\right]
\label{eq:ratio_final}
\end{equation}
\begin{figure}
    \begin{center}
\includegraphics[width=1.0\linewidth]{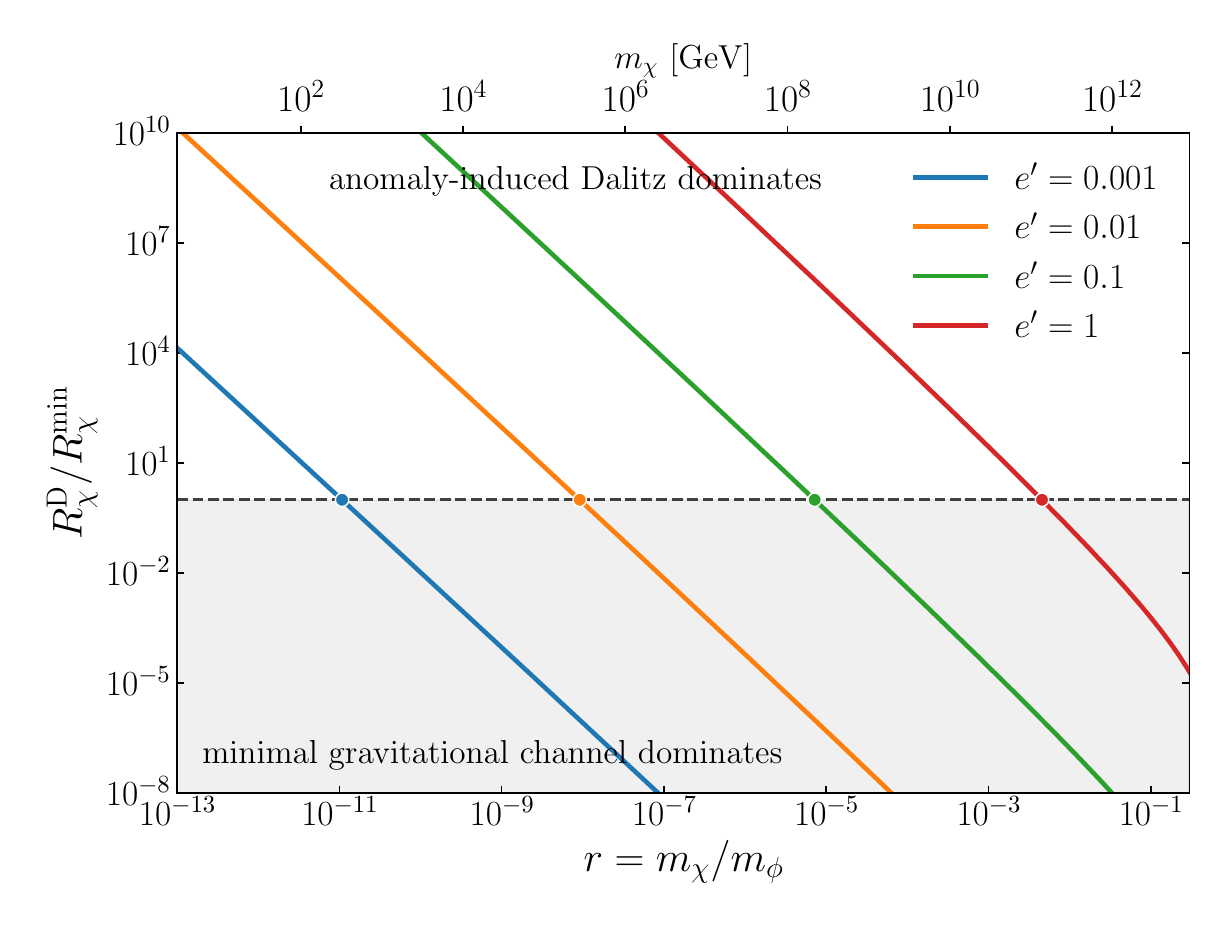}\end{center}
    \caption{\leftskip=0pt \rightskip=0pt \parfillskip=0pt plus 1fil Ratio of the anomaly-induced Dalitz production rate to the minimal gravitational fermion production rate as a function of $r=m_\chi/m_\phi$, for several values of the dark gauge coupling $e'$. The upper axis shows the corresponding dark fermion mass for fixed $m_\phi=3\times10^{13}\,\mathrm{GeV}$ and $N_\chi=1$; for $N_\chi \neq 1$, $e' \rightarrow N_\chi^{1/3} e'$, see Eq.\eqref{eq:ratio_final}.}
    \label{fig:ratio_rates}
\end{figure}
This is illustrated in Figure~\ref{fig:ratio_rates} where, for  $m_\phi=3\times10^{13}\,\mathrm{GeV}$, shows that the anomaly-induced
Dalitz channel becomes dominant below
$m_\chi\simeq3\times10^{2}$, $3\times10^{5}$,
$2\times10^{8}$ and $1.4\times10^{11}\,\mathrm{GeV}$
for $N_\chi^{1/3} e'=10^{-3}$, $10^{-2}$, $10^{-1}$ and $1$, respectively.\\

{
In addition to production from the inflaton condensate,
dark sector particles can be gravitationally produced
through scatterings in the SM thermal bath during reheating.
For dark fermions, the process,
\begin{equation*}
    \mathrm{SM}+\mathrm{SM}\to h^*\to\chi\bar{\chi}
\end{equation*}
gives the gravitational freeze-in rate
~\cite{Bernal:2018qlk,Mambrini:2021zpp,Clery:2021bwz},
\begin{equation}
    R_\chi^{\rm bath}
    = \kappa_\chi N_\chi \frac{T^8}{M_P^4},
    \label{eq:thermal_rate_chi}
\end{equation}
where $T$ denotes the temperature of the SM bath and
$\kappa_\chi=11351\pi^3/10368000\simeq0.034$.

Similarly, dark photons are produced through the process,
\begin{equation*}
  \mathrm{SM}+\mathrm{SM}\to h^*\to\gamma'\gamma'  
\end{equation*}
at a rate,
\begin{equation}
    R_{\gamma'}^{\rm bath}
    = \kappa_{\gamma'}\frac{T^8}{M_P^4},
    \label{eq:thermal_rate_gamma}
\end{equation}
with $\kappa_{\gamma'}\sim\kappa_\chi$
~\cite{Bernal:2018qlk}.
Unlike production from the inflaton condensate, this contribution is not suppressed in the massless dark
photon limit, since the thermal bath does not only probe the trace of the energy-momentum tensor, $T^{\m\n}$.
}

\subsection{Dark matter abundance}
\label{subsec:DM_abundance}

We now determine the relic abundance of dark fermions produced during reheating. We first assume that the produced dark sector particles remain out of equilibrium, and neglect sequential production. The validity of these assumptions will be discussed in the following subsections.\\

Denoting by $n_\chi$ the total number density of fermions and
antifermions, the Boltzmann equation reads
\begin{equation}
    \frac{dn_\chi}{dt}
    +3Hn_\chi
    =
    R_\chi,
    \label{eq:Boltzmann}
\end{equation}
where $R_\chi$ is the production rate for the relevant process. Introducing the comoving number density
\begin{equation}
    Y_\chi \equiv a^3 n_\chi ,
\end{equation}
Eq.\eqref{eq:Boltzmann} becomes
\begin{equation}
    \frac{dY_\chi}{da}
    =
    \frac{a^2}{H}\,
    R_{\chi}(a)
    \label{eq:Boltzmann_a}
\end{equation}
For a quadratic inflaton potential, neglecting the backreaction of
the radiation bath on the expansion rate, its energy density evolves as
\begin{equation}
    \rho_\phi(a)
    =
    \rho_{\rm end}
    \left(
        \frac{a_{\rm end}}{a}
    \right)^3 ,
    \label{eq:rho_phi_reheating}
\end{equation}
and consequently
\begin{equation}
    H(a)
    \simeq
    \sqrt{\frac{\rho_\phi(a)}{3M_P^2}}
    =
    H_{\rm end}
    \left(
        \frac{a_{\rm end}}{a}
    \right)^{3/2}
    \label{eq:H_reheating}
\end{equation}
until reheating is achieved when inflaton energy density reach SM radiation energy density, $\rho_{\phi}\equiv \rho_R$, defining the associated scale factor $a_{\rm RH}$.\\

We now focus on the anomaly-induced Dalitz source with the rate $R_\chi^{\rm D}$ derived in previous subsection. Since the anomaly-induced source scales as
$R_\chi^{\rm D}\propto\rho_\phi^2\propto a^{-6}$, production is dominated
by the earliest stages of the reheating phase. Integrating
Eq.~\eqref{eq:Boltzmann_a} from $a_{\rm end}$ to
$a_{\rm RH}$ gives
\begin{equation}
    n_\chi^{\rm D}(a_{\rm RH})
    =
    \frac{2}{3}
    \frac{R_\chi^{\rm D}(a_{\rm end})}{H_{\rm end}}
    \left(
        \frac{a_{\rm end}}{a_{\rm RH}}
    \right)^3
    \left[
        1-
        \left(
            \frac{a_{\rm end}}{a_{\rm RH}}
        \right)^{3/2}
    \right]
    \label{eq:nchi_general}
\end{equation}
For $a_{\rm RH}\gg a_{\rm end}$, we obtain, in the limit $m_\chi\ll m_\phi$,
\begin{equation}
   n_\chi^{\rm D}(a_{\rm RH})
    \simeq
    \frac{
        N_\chi^3 \alpha'^3 \rho_{\rm RH}\sqrt{\rho_{\rm end}}
    }{
        864\sqrt{3}\pi^4 M_P^3
    }
    \left[
        2\log\left(
            \frac{4m_\phi^2}{m_\chi^2}
        \right)-7
    \right]
    \label{Eq:nchi}
\end{equation}
As expected from the scaling $R_\chi^{\rm D}\propto a^{-6}$, the final
number density depends on the energy scale at the end of inflation
through $\sqrt{\rho_{\rm end}}$, whereas the subsequent dilution up
to reheating is encoded in $\rho_{\phi}(a_{\rm RH})\equiv \rho_{\rm RH}$.

After reheating, and in the absence of further dark sector number-changing
processes, the comoving dark-matter number density is conserved. We define the reheating temperature by
\begin{equation}
    \rho_{\rm RH}
    =
    \frac{\pi^2}{30}
    g_{*,{\rm RH}}
    T_{\rm RH}^4 
    \label{eq:rho_RH}
\end{equation}
Then, the
present relic abundance can therefore be written as \cite{Mambrini:2021zpp, Clery:2021bwz}
\begin{equation}
    \Omega_\chi h^2
    =
    1.6\times10^8\,
    \frac{g_{*s,0}}{g_{*s,{\rm RH}}}
    \frac{n_\chi(a_{\rm RH})}{T_{\rm RH}^3}
    \left(
        \frac{m_\chi}{1\,{\rm GeV}}
    \right),
    \label{eq:Omega_general}
\end{equation}
where $g_{*s,0}=3.91$ and we take $g_{*s,{\rm RH}}=106.75$ for reheating temperatures well above the electroweak scale.\\

Equations~\eqref{Eq:nchi} and \eqref{eq:Omega_general} show that the
anomaly-induced Dalitz contribution scales parametrically as
\begin{equation}
    \Omega_\chi^{\rm D}h^2
    \propto
    N_\chi^3 \alpha'^3\,
    m_\chi T_{\rm RH}\,
    \sqrt{\rho_{\rm end}}\,
    \left[
        2\log\left(
            \frac{4m_\phi^2}{m_\chi^2}
        \right)-7
    \right]
    \label{eq:Omega_scaling}
\end{equation}
which provides the observed DM relic \cite{Aghanim:2018eyx} for,
\begin{equation}
    \frac{\Omega_\chi^{\rm D}h^2}{0.12}
    \simeq 
    N_\chi^3\left(\frac{e'}{0.9}\right)^{6}
    \left(\frac{m_\chi}{10^{8}\,{\rm GeV}}\right)
    \left(\frac{T_{\rm RH}}{4\times 10^{13}\,{\rm GeV}}\right),
\end{equation}
up to the logarithmic dependence on $m_\chi/m_\phi$, here evaluated at $m_\chi = 10^8$ GeV, $m_\phi = 3\times10^{13}$ GeV, and $\rho_{\rm end}= (5.2\times 10^{15}\,\rm GeV)^4$, as for Starobinsky-like inflation \cite{Starobinsky:1980te}.
\\

Dark fermions are also produced through the minimal, leading-order gravitational portals. This includes production directly
from the inflaton condensate, whose production rate was given in Eq.~\eqref{eq:Rminimal}, as well
as production from the gravitational scattering of particles in the
visible thermal bath, at a rate given in Eq.(\ref{eq:thermal_rate_chi}). The latter is the standard gravitational freeze-in contribution from the thermal bath and has been computed in ~\cite{Mambrini:2021zpp, Clery:2021bwz}, In the following, we denote the corresponding number densities by $n_\chi^{\rm min}$, $n_\chi^{\rm bath}$, respectively. Integrating Eq.~(\ref{eq:Boltzmann_a}) between $a_{\rm end}$ and $a_{\rm RH}$, one finds \cite{Clery:2021bwz},
\begin{equation}
n^{\rm min}_\chi(a_{\rm RH})\simeq\frac{N_\chi\,\rho_{\rm RH}\sqrt{\rho_{\rm end}}}{64\sqrt3\,\pi M_P^3} \left(\frac{m_\chi}{m_\phi}\right)^2\left(1-\left(\frac{m_\chi}{m_\phi}\right)^2\right)^{3/2}
\end{equation}
and,
\begin{equation}
n^{\rm bath}_\chi(a_{\rm RH})\simeq\frac{2\kappa_\chi N_\chi}{\sqrt3\,\alpha_R^{1/2}}\frac{T_{\rm RH}^6}{M_P^3},
\end{equation}
with $\alpha_R=\pi^2g_{*,{\rm RH}}/30$. They give the following DM relics \cite{Clery:2021bwz},
\begin{eqnarray}
\frac{\Omega^{\rm min}_\chi h^2}{0.12}&\simeq&N_\chi
\left(\frac{m_\chi}{10^{10}\,{\rm GeV}}\right)^3
\frac{T_{\rm RH}}{10^{14}\,{\rm GeV}}
\label{eq:min}
\end{eqnarray}
and
\begin{equation}
\frac{\Omega^{\rm bath}_\chi h^2}{0.12}\simeq
N_\chi\frac{m_\chi}{10^{10}\,{\rm GeV}}
\left(\frac{T_{\rm RH}}{10^{13}\,{\rm GeV}}\right)^3 
\label{eq:bath}
\end{equation}
for $m_\phi = 3\times 10^{13}\, \rm GeV$ and $\rho_{\rm end}=(5.2\times10^{15}\, \rm GeV)^4$

Together with the anomaly-induced Dalitz contribution, we include
these populations  when determining the total dark-matter
abundance and, below, when estimating the efficiency of dark sector
thermalization. In the absence of significant dark sector interactions, the total fermion abundance from gravitational portals is therefore obtained from
\begin{equation}
    \Omega_\chi^{\rm tot}h^2
    =
    \Omega_\chi^{\rm D}h^2
    +
    \Omega_\chi^{\rm min}h^2
    +
    \Omega_\chi^{\rm bath}h^2 
\end{equation}
This ceases to be appropriate if the dark sector interactions become efficient, eventually leading to its thermalization. We discuss the thermalization condition in the section \ref{subsec:thermalization}.

\subsection{Transverse dark photon abundance}

The anomaly-induced gravitational portal also produces a population of dark photons with transverse polarization. Their number density obeys
\begin{equation}
    \frac{dn_{\gamma'_T}}{dt}
    +3Hn_{\gamma'_T}
    =
    R_{\gamma'_T} 
\end{equation}
with $R_{\gamma'_T}$ given in Eq.(\ref{eq:Rgamma}). 
Since $R_{\gamma'_T}\propto \rho_\phi^2\propto a^{-6}$, the integration
proceeds exactly as for the Dalitz-produced fermions. During the
matter-dominated reheating stage, we find
\begin{equation}
    n_{\gamma'_T}(a)
    =
    \frac{2}{3}
    \frac{R_{\gamma'_T}(a_{\rm end})}{H_{\rm end}}
    \left(
        \frac{a_{\rm end}}{a}
    \right)^3
    \left[
        1-
        \left(
            \frac{a_{\rm end}}{a}
        \right)^{3/2}
    \right]
    \label{eq:ngamma}
\end{equation}
At the end of reheating, for $a_{\rm RH}\gg a_{\rm end}$,
\begin{equation}
    n_{\gamma'_T}(a_{\rm RH})
    \simeq
    \frac{
        N_\chi^2 \alpha'^2\rho_{\rm RH}\sqrt{\rho_{\rm end}}
    }{
        288\sqrt{3}\pi^3M_P^3
    }
    \label{eq:ngamma_RH}
\end{equation}
Unsurprisingly, the number density of dark photons is parametrically larger than the one of dark electrons produced through the Dalitz process. Since  dark photons are produced with energy $E_{\gamma'}\simeq m_\phi$, their energy density satisfies
\begin{equation}
    \rho_{\gamma'_T}(a_{\rm RH})
    \simeq
    \frac{N_\chi^2\alpha'^2}{96\sqrt{3}\,\pi^3}\,
    \frac{
        m_\phi\,
        \rho_{\rm end}^{1/6}\,
        \rho_{\rm RH}^{4/3}
    }{
        M_P^3
    }
    \label{eq:rho_gamma_RH}
\end{equation}
and
\begin{equation}
    \frac{\rho_{\gamma'_T}(a_{\rm RH})}{\rho_{\rm RH}}
    \simeq
   \frac{N_\chi^2\alpha'^2}{96\sqrt{3}\,\pi^3}
    \frac{m_\phi}{M_P^3}
    \rho_{\rm end}^{1/6}\rho_{\rm RH}^{1/3}
    \ll 1 
    \label{eq:dark_radiation_fraction}
\end{equation}

{For the dark photons produced gravitationally from the SM thermal bath, at a rate given in Eq.(\ref{eq:thermal_rate_gamma}), their number density at reheating, is given by \cite{Clery:2021bwz},
\begin{equation}
n^{\rm bath}_{\gamma'}(a_{\rm RH})\simeq\frac{2\kappa_{\gamma'}}{\sqrt3\,\alpha_R^{1/2}}\frac{T_{\rm RH}^6}{M_P^3},
\end{equation}
which, relative to Eq.~(\ref{eq:ngamma_RH}), is then,
\begin{eqnarray}
\frac{n^{\rm bath}_{\gamma'}}{n_{\gamma'_T}}&\simeq&\frac{576\pi^3\kappa_{\gamma'}}{N_\chi^2\alpha'^2\alpha_R^{3/2}}\frac{T_{\rm RH}^2}{\sqrt{\rho_{\rm end}}}\nonumber \\
&\sim& 0.1\left(\frac{0.08}{\alpha'}\right)^2\left(\frac{T_{\rm RH}}{10^{14}\,{\rm GeV}}\right)^2
\label{eq:ngamma_bath}
\end{eqnarray}
They carry an energy density, $\rho^{\rm bath}_{\gamma'}(a_{\rm RH})/\rho_{\rm RH}\sim10^{-17}$ for $T_{\rm RH}\sim10^{14}$~GeV. Their contribution to dark radiation is thus completely negligible.}\\

The produced dark photons constitute thus only a negligible fraction of the total radiation energy density. Their density can however be relevant for subsequent interactions within the
dark sector. Also, at least in principle, these dark photons could subsequently annihilate into fermion-antifermion pairs, $
    \gamma'\gamma' \rightarrow \chi\bar{\chi}$
and thereby provide an additional source of DM. However, we show in Appendix \ref{appendix:secondary} that this contribution is negligible.

\subsection{Dark photon mass}
\label{subsec:DP_mass}
So far, we have taken the dark photon to be strictly massless. Yet, if $\chi$ constitutes the DM, a massless dark photon mediates a long-range force between DM particles. Constraints from binary self-scattering are weak for the large DM masses considered here~\cite{Ackerman:2008kmp, Feng:2009mn,Agrawal:2016quu,Chu:2011be}, but collective plasma instabilities in colliding clusters are much more constraining~\cite{Lasenby:2020rlf}. For an exactly massless mediator, particle-in-cell simulations of the Bullet Cluster give \cite{DeRocco:2024ifs}, $
e'\lesssim3\times10^{-14}\,(m_\chi/{\rm GeV})$, which excludes the coupling and mass ranges considered in this work. These instabilities develop on time scales set by the dark plasma frequency~\cite{Lasenby:2020rlf},
\begin{eqnarray}
\omega_p=e'\frac{\sqrt{\rho_\chi}}{m_\chi}&\simeq&4\times10^{-23}\,{\rm eV}\left(\frac{e'}{0.5}\right)\\
\nonumber
&\times&\left(\frac{10^{10}\,{\rm GeV}}{m_\chi}\right)\left(\frac{\rho_\chi}{0.1\,{\rm GeV\,cm^{-3}}}\right)^{1/2},
\label{eq:omegap}
\end{eqnarray}
where the local DM density $\rho_\chi$ is normalized to a typical value for colliding clusters. 
Thus, a dark photon mediator mass $m_{\gamma'}\gtrsim\omega_p/\sigma$, with $\sigma\sim10^3$~km/s the DM velocity dispersion, suppresses these instabilities, while purely magnetic instabilities are already suppressed for $m_{\gamma'}\gtrsim\mathcal{O}(\omega_p)$~\cite{Lasenby:2020rlf}. A dark photon mass as low as $m_{\gamma'}\gtrsim10^{-20}\, \rm eV$ is therefore expected to evade these constraints. The precise lower mass required to make the collective instabilities is somewhat system-dependent. Here, we do not impose a sharp bound, merely noting that a tiny dark photon mass would be sufficient to make such effects harmless.

However, even a small dark photon mass has an impact on gravitational particle production, as a massive dark photon has an additional longitudinal polarization, $\gamma'_L$. For energies $E\gg m_{\gamma'}$, it behaves as the Goldstone mode eaten by the dark photon and thus, for gravitational interaction, as a minimally coupled scalar. Concretely, its stress-energy tensor is not traceless and so its gravitational production does not vanish as $m_{\gamma'}$ goes to zero ~\cite{Kolb:2020fwh,Garcia:2023obw}. We emphasize first that such dark photon longitudinal modes production does not affect the production of dark fermions discussed so far: since the dark current is conserved, the $k^\mu/m_{\gamma'}$ part of the longitudinal polarization vector drops out, and each longitudinal leg coupled to $\chi$ costs a factor $m_{\gamma'}/E$ in the amplitude. For $m_{\gamma'}\ll m_\chi$  the Dalitz rate, the production of transverse dark photons and the sequential production (see Appendix~\ref{appendix:secondary}) are unchanged, up to corrections $\mathcal{O}(m_{\gamma'}^2/m_\chi^2)$. As we will indeed consider very light dark photons, their contribution to this process is totally negligible. Still, it remains to be discussed whether the longitudinal population itself is affecting DM  or dark radiation relics.

Indeed, the longitudinal mode of massive dark photons can be produced during inflation~\cite{Graham:2015rva,Kolb:2020fwh}. Modes with comoving momentum $k/a\gg m_{\gamma'}$ behave again as a light Goldstone scalar and acquire fluctuations of order $H_I/2\pi$ when they exit the horizon. After inflation, short-wavelength modes re-enter the horizon while still relativistic and redshift as radiation, while the energy density in super-horizon modes redshifts as $a^{-2}$ as long as $H>m_{\gamma'}$, suppressing power on large scales~\cite{Graham:2015rva}. The resulting energy spectrum is therefore peaked at modes re-entering the horizon when $H\simeq m_{\gamma'}$. These have $k/a\simeq m_{\gamma'}$, and are non-relativistic as soon as they start to oscillate. This population is then cold and behaves as DM, with an abundance~\cite{Graham:2015rva}
\begin{equation}
\frac{\Omega^{\rm inf}_{\gamma'_L}}{\Omega_{\rm DM}}\simeq\left(\frac{m_{\gamma'}}{6\times10^{-6}\,{\rm eV}}\right)^{1/2}\left(\frac{H_I}{10^{14}\,{\rm GeV}}\right)^2 
\label{eq:GMR}
\end{equation}
where $H_I$ is the inflationary Hubble scale.
Since $H\simeq m_{\gamma'}$ is reached long after reheating, this estimate does not depend on the details of the reheating phase. 

Seemingly, during reheating, $\gamma'_L$ modes are produced through the inflaton gravitational portal, $\phi\phi\to h^*\to\gamma'_L\gamma'_L$, again with no mass suppression. In the limit $m_{\gamma'}\ll m_\phi$, one finds~\cite{Garcia:2023obw}
\begin{equation}
n_{\gamma'_L}(a_{\rm RH})\simeq\frac{\rho_{\rm RH}\sqrt{\rho_{\rm end}}}{64\sqrt3\,\pi M_P^3},
\label{eq:nL} \implies \frac{n_{\gamma'_L}}{n_{\gamma'_T}}=9\pi^2/(2N_\chi^2\alpha'^2)>1
\end{equation}
As expected, longitudinal modes outnumber the transverse ones, whose production via the trace anomaly is loop-suppressed. There is also production from the SM thermal bath, $\mathrm{SM}+\mathrm{SM}\to h^*\to\gamma'_L\gamma'_L$, but it is much subdominant compared to the one from the inflaton condensate for reasonable reheating temperatures ~\cite{Clery:2021bwz}. These longitudinal dark photons are produced with energy $E= m_\phi$, and thus form a population of dark radiation. The hot population produced during reheating carries a negligible energy density, $\rho_{\gamma'_L}/\rho_{\rm RH}\lesssim10^{-14}$ for $T_{\rm RH}\lesssim10^{14}$~GeV, and the transverse population is even smaller, see Eq.~(\ref{eq:dark_radiation_fraction}). Their contribution to dark radiation, $\Delta N_{\rm eff}$, is thus completely negligible. 
\\

In the absence of kinetic mixing and for $m_{\gamma'}<2m_\chi$, the dark photon is stable\footnote{An unstable dark photon, e.g.\ for $m_{\gamma'}>2m_\chi$ or through kinetic mixing, $\epsilon$, with the SM, or a more extended dark sector, would lead to a different phenomenology, which is beyond the scope of this work.}, and both populations eventually contribute to the late DM density. Requiring that the population produced during reheating does not exceed the dark fermions produced through the Dalitz process gives
\begin{eqnarray}
\frac{m_{\gamma'}}{m_\chi}\lesssim\frac{n^{\rm D}_\chi}{n_{\gamma'_L}}&\simeq&\frac{2N_\chi^3\alpha'^3}{27\pi^3}\left[2\log\left(\frac{4m_\phi^2}{m_\chi^2}\right)-7\right]\nonumber\\
&\sim& 10^{-6}\left(\frac{e'}{0.5}\right)^6,
\label{eq:mLreh}
\end{eqnarray}
The cold population of dark photons produced during inflation is much more constraining. According to Eq.~\eqref{eq:GMR}, it remains subdominant provided
\begin{equation}
m_{\gamma'}\lesssim6\times10^{-6}\,{\rm eV}\left(\frac{10^{14}\,{\rm GeV}}{H_I}\right)^4 ,
\label{eq:mLinf}
\end{equation}
where $H_I$ should be evaluated when the modes $k$ exit the horizon. For Starobinsky-like inflation, $H_I\simeq
m_\phi/2\simeq1.5\times10^{13}$~GeV, and the inflationary population remains subdominant for $m_{\gamma'}\lesssim10^{-2}$~eV. 

In summary, we conclude that a stable, massive dark photon with mass in the range,
\begin{equation}
10^{-20}\,{\rm eV}\lesssim m_{\gamma'}\lesssim10^{-2}\,{\rm eV}
\end{equation}
evades the constraints. It has a longitudinal modes population that is negligible both as dark radiation and as DM, leaving our results on dark fermion production through the trace anomaly unchanged.
\\

These constraints still depend on how the dark photon mass is generated. For a Stueckelberg mass, $\gamma'_L$ exists at all times, and both populations are produced, the inflationary one setting the strongest bound~\eqref{eq:mLinf}. If instead the mass were to arise from a dark Higgs mechanism, the vacuum expectation value $v'$ is tiny, and the $U(1)'$ symmetry is generically unbroken during inflation and reheating. The dark photon is then massless at these epochs, and there is no inflationary production of $\gamma'_L$, so that Eq.~\eqref{eq:mLinf} does not apply~\cite{Graham:2015rva}. The light charged dark Higgs is, however, itself produced gravitationally, its phase becoming $\gamma'_L$ after symmetry breaking, so that Eq.~\eqref{eq:mLreh} remains a relevant estimate up to an $\mathcal{O}(1)$ factor. It also contributes to $\beta(e')$, and thus to the anomalous production. In both cases, a mass generated or present only at late times, after reheating, would leave the DM production studied here essentially unchanged.

\subsection{Dark sector thermalization}
\label{subsec:thermalization}

The relic abundance derived in section~\ref{subsec:DM_abundance}
assumes that the gravitationally populated dark sector never reaches internal thermal equilibrium. We will now check the validity of this assumption. The thermalization of a gravitationally produced dark QED sector was studied in~\cite{Garny:2018grs} for gravitational production from the SM thermal bath. Here, the dark sector is in addition populated by the inflaton condensate, through the minimal and anomaly-induced channels.\\

Assuming an under-occupied, relativistic dark sector, its thermalization proceeds through a cascade of collinear splittings of particles, a process which is generically affected by the so-called Landau-Pomeranchuk-Migdal (LPM) effect~\cite{Kurkela:2011ti,Harigaya:2013vwa,Garny:2018grs,Clery:2026fhj}. The characteristic
thermalization rate can
be estimated as being of order\footnote{{A binary collision exchanges one dark photon between two DM particles, so its cross section is $e'^4$ while there are $N_\chi$ scatterers. For fixed $N_\chi \alpha'$, thermalization is $1/N_\chi$ suppressed \cite{Moore:2001fga}.}}
\begin{equation}
    \Gamma_{\rm th}
    \simeq
   N_\chi \alpha'^2
    \left[
        \sum_i \frac{n_i}{E_i}
    \right]^{1/2},
\label{eq:Gamma_thermalization}
\end{equation}
where $n_i$ and $E_i$ denote respectively the number density and
typical energy of the population of hard or comparatively high energy particles. In the present scenario, this includes only the $N_\chi$ charged particles,
\begin{equation}
    i=
    \chi^{\rm D},
    \quad
    \chi^{\rm min},
    \quad
    \chi^{\rm bath},
    \label{eq:thermal_populations}
\end{equation}
corresponding respectively to
Dalitz produced fermions, minimally gravitationally produced
fermions from the inflaton condensate, and fermions gravitationally produced from the visible thermal bath \cite{Clery:2021bwz}. The dark photons themselves do not contribute; being neutral, they do not screen nor broaden the momentum of hard particles, and only undergo rare hard scatterings, see \cite{Clery:2026fhj}. 

Since production from the inflaton condensate is UV dominated, these
particles are predominantly generated near the beginning of the
reheating phase. Their characteristic energy at the end of reheating is therefore redshifted to
\begin{equation}
    E_i(a_{\rm RH})
    \simeq
    m_\phi
    \left(
        \frac{H_{\rm RH}}{H_{\rm end}}
    \right)^{2/3}
    \label{eq:E_RH_condensate}
\end{equation}
For fermions produced from the visible thermal bath, we instead take
\begin{equation}
    E_\chi^{\rm bath}(a_{\rm RH})
    \simeq
    T_{\rm RH}
    \label{eq:E_RH_bath}
\end{equation}
We consider the thermalization parameter at reheating as
\begin{equation}
    \left.
    \frac{\Gamma_{\rm th}}{H}
    \right|_{\rm RH}
    =
    \frac{N_\chi \alpha'^2}{H_{\rm RH}}
    \left[
        \frac{n_\chi^{\rm D}}{E_\chi^{\rm D}}
        +
        \frac{n_\chi^{\rm min}}{E_\chi^{\rm min}}
        +
        \frac{n_\chi^{\rm bath}}{E_\chi^{\rm bath}}
    \right]_{\rm RH}^{1/2}
    \label{eq:X_RH}
\end{equation}
If
\begin{equation}
    \left.
    \frac{\Gamma_{\rm th}}{H}
    \right|_{\rm RH}\gtrsim1,
\end{equation}
dark sector interactions become efficient before or around the end
of reheating. In this region the abundance obtained
in section~\ref{subsec:DM_abundance} is no longer applicable.
Even when the criterion is not satisfied during reheating, thermalization of the DS may occur later on. As long as the dark sector particles remain relativistic
and decoupled from the visible bath, after reheating
\begin{equation}
    n_i\propto a^{-3},
    \qquad
    E_i\propto a^{-1},
\end{equation}
such that
\begin{equation}
    \frac{\Gamma_{\rm th}}{H}\propto a
\end{equation}
during radiation domination.

\begin{figure}
    \centering
    \includegraphics[width=1.0\linewidth]{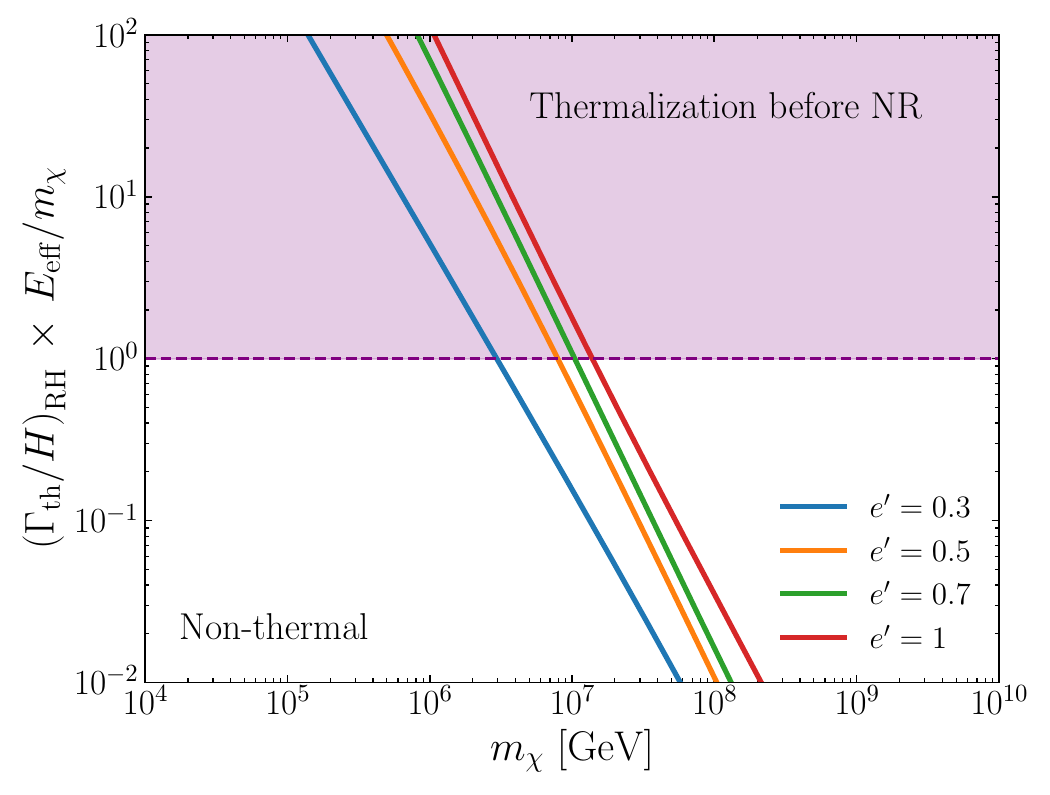}
    \caption{\leftskip=0pt \rightskip=0pt \parfillskip=0pt plus 1fil Thermalization parameter $(\Gamma_{\rm th}/H)_{\rm RH}\times E_{\rm eff}/m_\chi$, evaluated along the $\Omega_{\rm tot}h^2=0.12$ contour, for several values of the dark gauge coupling $e'$. The horizontal dashed line marks the condition for thermalization before the would-be non-relativistic transition. We take $N_\chi=1$, $m_\phi=3\times10^{13}\,\mathrm{GeV}$ and $\rho_{\rm end}^{1/4}=5.2\times10^{15}\,\mathrm{GeV}$.}
    \label{fig:thermalization_relic}
\end{figure}

To estimate whether thermalization occurs before the fermions become
non-relativistic, it is convenient to introduce the effective energy
\begin{equation}
    E_{\rm eff}^{-1}
    \equiv
    {
        \displaystyle\sum_i n_i/E_i
    }/{
        \displaystyle\sum_i n_i
    }
    \label{eq:E_eff}
\end{equation}
Then, the dark sector can thermalize while the fermions are still
relativistic provided 
\begin{equation}
    \left.
    \frac{\Gamma_{\rm th}}{H}
    \right|_{\rm RH}
    \gtrsim
    \frac{m_\chi}{E_{\rm eff}(a_{\rm RH})}
    \label{eq:thermalization_boundary}
\end{equation}
approximately.
To illustrate the relevance of this condition for non-thermal
scenarios, we show in figure~\ref{fig:thermalization_relic} 
$(\Gamma_{\rm th}/H)_{\rm RH}\times E_{\rm eff}/m_\chi$ evaluated along the contour
$\Omega_{\rm tot}h^2=0.12$. Values above unity correspond to
thermalization before the non-relativistic transition,
whereas values below unity remain within the non-thermal regime.
We use Eq.~\eqref{eq:thermalization_boundary} to identify the
region of parameter space in which the non-thermal freeze-in
calculation ceases to be reliable and the dark sector reaches eventually thermal equilibrium. If this is the case, DM abundance is set by thermal freeze-out \cite{Arvanitaki:2021qlj,Clery:2026fhj}. \\

\begin{figure*}[t]
  \centering
  \includegraphics[width=0.48\textwidth]{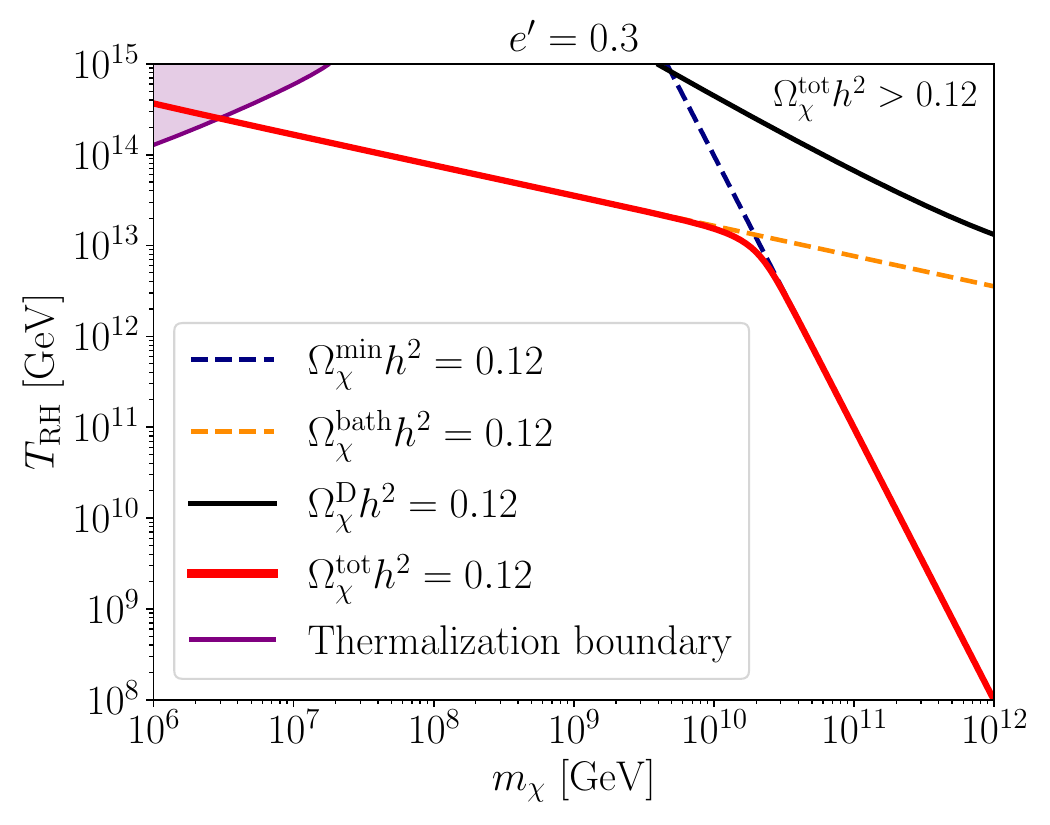}\hfill
  \includegraphics[width=0.48\textwidth]{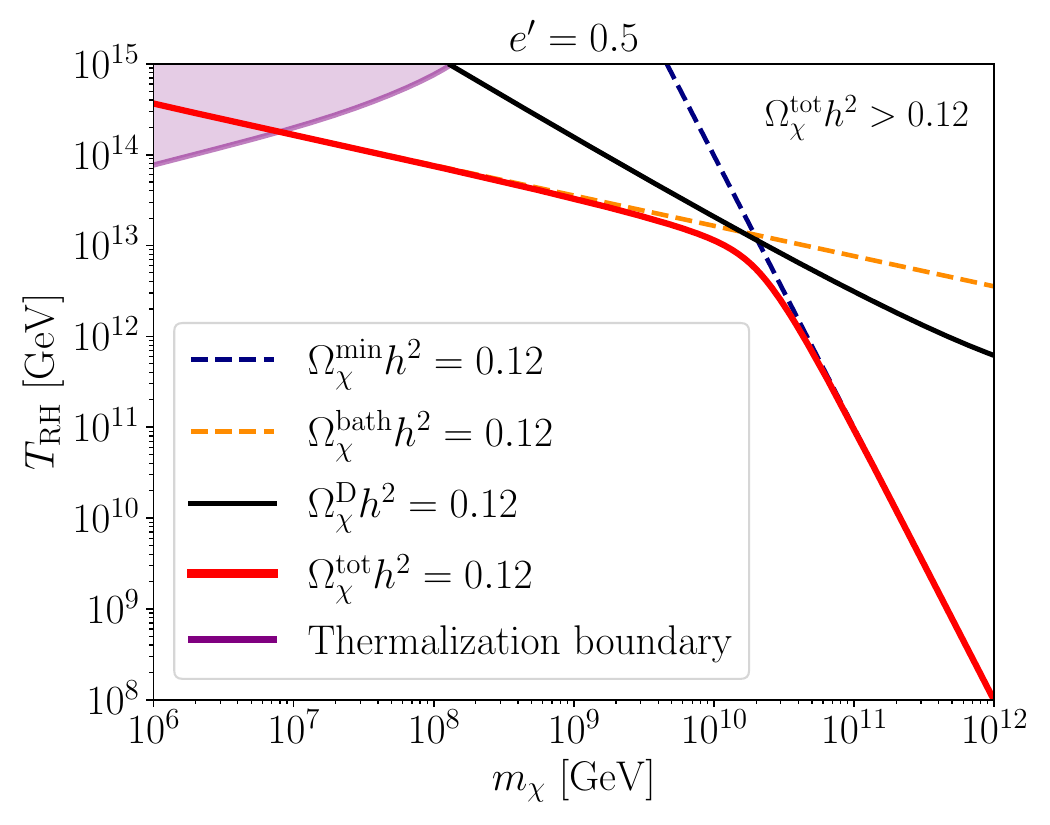}\\[2mm]
  \includegraphics[width=0.48\textwidth]{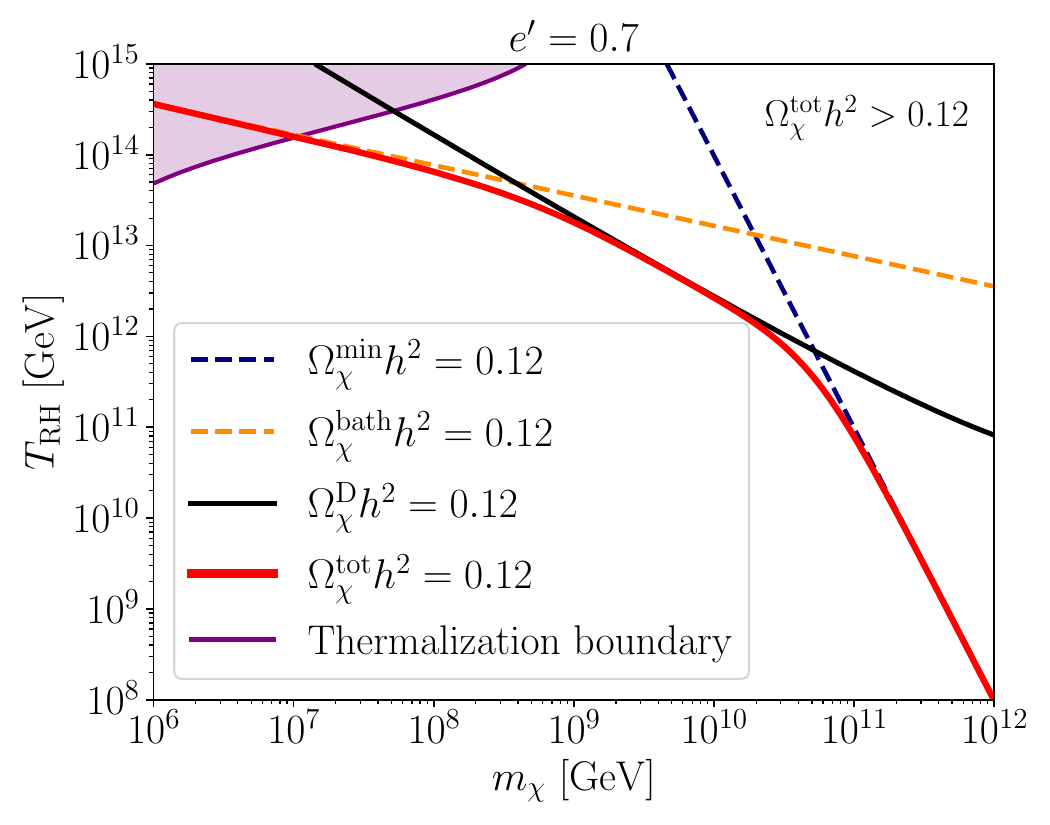}\hfill
  \includegraphics[width=0.48\textwidth]{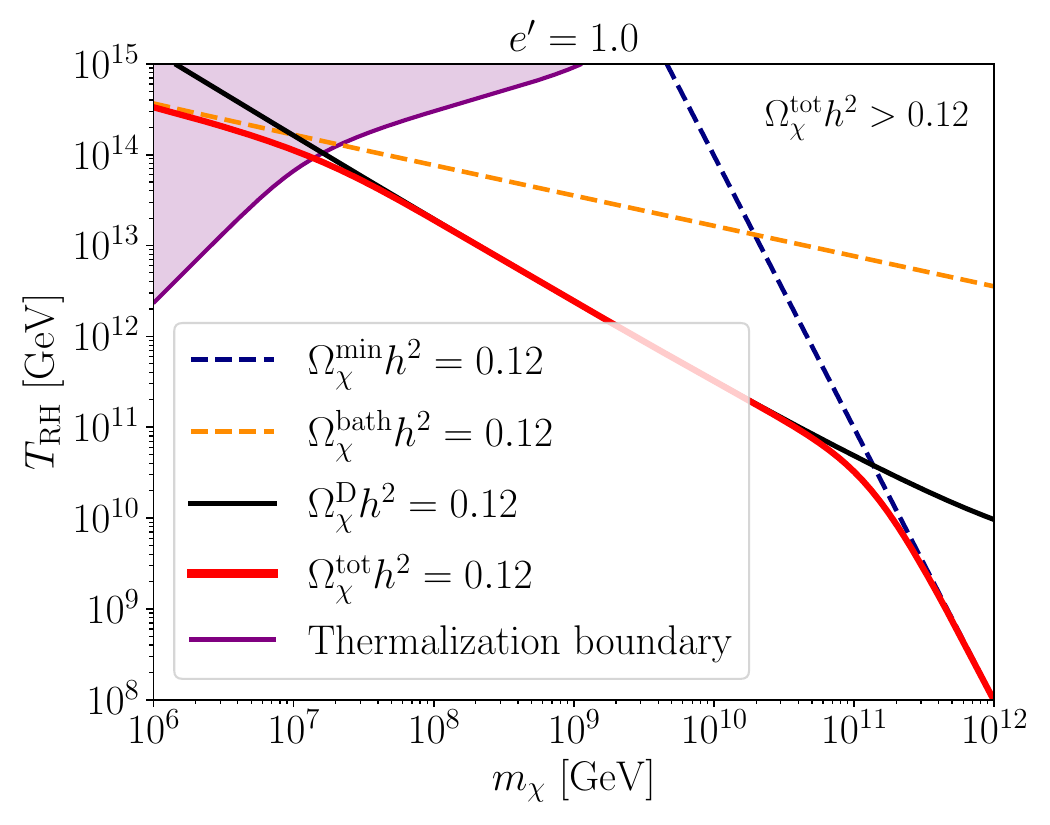}
  \caption{\leftskip=0pt \rightskip=0pt \parfillskip=0pt plus 1fil Contours of $\Omega_\chi h^2=0.12$ in the $(m_\chi,T_{\rm RH})$ plane for
  $e'=0.3$ (top left), $0.5$ (top right), $0.7$ (bottom left) and $1$ (bottom right):
  minimal production from the inflaton condensate (blue dashed), from the SM thermal
  bath (orange dashed), anomaly-induced Dalitz channel (black) and total (red). In the
  purple region, the dark sector thermalizes before $\chi$ becomes non-relativistic,
  Eq.~(\ref{eq:thermalization_boundary}), and the non-thermal result does not apply. We take $N_\chi=1$,
  $m_\phi=3\times10^{13}\, \rm GeV$ and $\rho_{\rm end}^{1/4}=5.2\times10^{15}\, \rm GeV$.}
  \label{fig:relics}
\end{figure*}

\subsection{Comparison of DM production channels}

In figure~\ref{fig:relics}, we show the contours $\Omega_\chi h^2=0.12$ in the
$(m_\chi,T_{\rm RH})$ plane, varying the dark gauge coupling in the different panels, $N_\chi^{1/2} e'=0.3,\,0.5,\,0.7$ and $1$. We show the contours for the minimal gravitational production from the inflaton condensate (in dashed blue), the gravitational production from the SM thermal bath (dashed orange), the anomaly-induced Dalitz channel (solid black), and their sum (solid red). 

The three channels scale differently,
$\Omega^{\rm min}_\chi\propto m_\chi^3T_{\rm RH}$, $\Omega^{\rm bath}_\chi\propto m_\chi T_{\rm RH}^3$ and $\Omega^{\rm D}_\chi\propto \alpha'^3 m_\chi T_{\rm RH}$ (up to logarithms). The minimal channel dominates at large masses $m_\chi \gtrsim 10^{10}-10^{11}\, \rm GeV$, while the thermal bath contribution dominates at the highest reheating temperatures. The anomaly induced channel can only be dominant for lower $m_\chi$ and lower $T_{\rm RH}$. With a $(N_\chi \alpha')^3$ scaling, the outcome depends strongly on the dark gauge
coupling. Setting $N_\chi = 1$, for $e'=0.3$ the total abundance follows the minimal channels everywhere.
For $e'=0.5$ the Dalitz channel dominates in a narrow window around $m_\chi\sim10^{10}\, \rm GeV$. For $e'=0.7$ it dominates for $2\times10^{8}\lesssim m_\chi\lesssim5\times10^{10}\, \rm GeV$, and for $e'=1$ for $10^{7}\lesssim m_\chi\lesssim10^{11}\, \rm GeV$. In this range, the DM can thus be much lighter at a fixed reheating temperature than in the non-interacting HS scenario: for $e'=1$, the required mass drops from
$4.6\times10^{10}$ to $2.7\times10^{9}\, \rm GeV$ at $T_{\rm RH}=10^{12}\, \rm GeV$, and from $1.8\times10^{10}$ to $2\times10^{8}\, \rm GeV$ at $T_{\rm RH}=10^{13}\, \rm GeV$. As for all gravitational production channels, a rather large reheating temperature is still required, $T_{\rm RH}\gtrsim10^{10}\, \rm GeV$.

The purple regions show where the dark sector thermalizes before $\chi$ becomes
non-relativistic, see Eq.~(\ref{eq:thermalization_boundary}), so that the non-thermal production no longer provides the DM abundance.
Along the $\Omega^{\rm tot}_\chi h^2=0.12$ contours this happens below $m_\chi\simeq3\times10^{6},\,8\times10^{6},\,1.1\times10^{7}$ and $1.4\times10^{7}\, \rm GeV$
for $e'=0.3,\,0.5,\,0.7$ and $1$ (see figure \ref{fig:thermalization_relic}), up to $\mathcal O(1)$ factors. For $e'=1$, the anomaly-dominated window therefore extends down to $m_\chi\sim10^{7}\, \rm GeV$, for
$T_{\rm RH}\sim10^{14}\, \rm GeV$. At lower masses the dark sector thermalizes and the relic abundance is set by dark sector freeze-out during or after reheating, as discussed in \cite{Clery:2026fhj}.\footnote{{It is easy to account for a $N_\chi$ factor taking into account that the anomaly-induced channel scales as $N_\chi^3 \alpha'^3$ while the classical, helicity suppressed channel scales and thermalization rate both scale as $N_\chi$, see  Eqs. \eqref{eq:Rdalitz} and \eqref{eq:Gamma_thermalization}.  }}\\

If the dark sector thermalizes, it does so at a temperature $T'$ lower than that of the SM, $T$. Since, we consider no sizable portal interaction between the HS and the SM via the kinetic mixing $\epsilon$, the two sectors do not exchange energy further, and $T'$ follows from energy conservation within the dark sector. Its energy density is then shared between dark photons and dark fermions, $\rho'=\frac{\pi^2}{30}g'_*T'^4$ with $g'_*=2+N_\chi\times\frac78\times4 =5.5$ for $N_\chi = 1$. This gives at the end of reheating, 
\begin{eqnarray}
\xi_{\rm RH} &\equiv& \left({T'\over T}\right)_{\rm RH}\\
&\simeq&\left(\frac{g_*}{g'_*}\right)^{\frac14}\,\left[\frac{\rho_{\gamma'_T}+\rho_{\gamma'}^{\rm bath}+\rho^{\rm D}_\chi+\rho^{\rm min}_\chi+\rho^{\rm bath}_\chi}{\rho_{\rm RH}}\right]^{\frac14}\nonumber
\end{eqnarray}
or

\begin{eqnarray}
\label{eq:Tprime}
\xi_{\rm RH}
&\simeq&\left(\frac{g_*}{g'_*}\right)^{\frac14}\left[\frac{N_\chi^2\alpha'^2}{96\sqrt3\,\pi^3}\right.\\
&\times&\left(1+\frac{N_\chi\alpha'}{3\pi}\left(2\log\left(\frac{4m_\phi^2}{m_\chi^2}\right)-7\right)\right)\left.\frac{m_\phi\rho_{\rm end}^{1/6}\rho_{\rm RH}^{1/3}}{M_P^3}\right.
\nonumber\\
&+&\left.\frac{2(\kappa_\chi N_\chi+ \kappa_{\gamma'})}{\sqrt3\,\alpha_R^{3/2}}\frac{T_{\rm RH}^3}{M_P^3}\right]^{\frac14}\nonumber
\end{eqnarray}
where $g_*=106.75$ is the number of degrees of freedom in SM bath at the temperature of reheating. Here we have not included the longitudinal dark photons discussed in section~\ref{subsec:DP_mass}, which decouple from the dark current and therefore do not take part in the thermalization of the dark sector. {Since gravitational production is irreducible, Eq.~\eqref{eq:Tprime} provides a lower bound on the temperature of a dark QED sector for a given reheating temperature of the visible sector, almost independently of $m_\chi$. At large reheating temperatures it is dominated by the production from the SM thermal bath, $\xi_{\rm RH}\gtrsim 10^{-4}\,(T_{\rm RH}/10^{14}\,{\rm GeV})^{3/4}$, consistent with~\cite{Garny:2018grs}, while at lower reheating temperatures the anomaly-induced production from the inflaton condensate dominates, $\xi_{\rm RH}\gtrsim10^{-5}\,(\alpha'/0.08)^{1/2}\,(T_{\rm RH}/10^{12}\,{\rm GeV})^{1/3}$, both for $N_\chi=1$, $m_\phi=3\times10^{13}$~GeV and $\rho_{\rm end}=(5.2\times10^{15}\,{\rm GeV})^4$. This ratio remains constant after reheating as long as both sectors are relativistic and $g_*$ does not change. The domain of thermal candidates of a hidden sector that consists of dark QED was considered in \cite{Coy:2021ann,Coy:2024itg,Clery:2026fhj}}. The DM abundance in such a cold hidden sector is set by secluded freeze-out of $\chi\bar\chi\to\gamma'\gamma'$ at $T'$~\cite{Pospelov:2007mp,Chu:2011be}. This, in principle, opens the possibility of extending the range of viable dark matter produced through the gravitational portal to masses even lower than  $m_\chi \sim 10^7$ GeV. We leave this for future work.

\section{Conclusions}
In this work, we have investigated the extent to which the gauge interaction of
a dark QED sector can affect the gravitational production of DM 
during reheating after inflation. As is well known, the minimal gravitational production of light
fermions from the inflaton condensate is helicity suppressed. Such suppression is expected on very general grounds and occurs for many DM portals. However, for the graviton portal, the emission of an additional dark photon 
is not sufficient to lift the helicity suppression because the classical
stress-energy tensor of dark QED is traceless in the conformal limit, $T^\mu_\mu \rightarrow 0$. We found it a useful exercise to check explicitly the associated Ward identity for the process of (dark) photon bremsstrahlung. 

At one-loop order, the
trace anomaly generates an effective coupling of the graviton to massless
dark photons, providing an irreducible source of dark photon radiation, albeit suppressed by powers of the dark fine structure constant. The anomalous channel opens the possibility of producing dark fermions through an off-shell dark photon (akin to Dalitz decay of a neutral pion, $\pi^0 \rightarrow \gamma e^+e^-$), a process which is enhanced by a factor of $\log(m_{\phi}/m_\chi)$. We have shown that the anomalous channel can dominate over the minimal gravitational
production of fermions, provided the dark fermion is sufficiently light and the dark gauge coupling sufficiently large. {For $e'\simeq1$, the DM mass required at fixed reheating temperature is reduced by one to two orders of magnitude, and the viable mass range extends down to $m_\chi\sim10^7$~GeV for $T_{\rm RH}\sim10^{14}$~GeV.}

We have examined the evolution of the gravitationally
produced dark QED hidden sector, paying in particular attention to the different production channels, which affect differently the longitudinal and transverse modes of the dark photon.
{A small dark photon mass, $10^{-20}\,{\rm eV}\lesssim m_{\gamma'}\lesssim10^{-2}$~eV for a Stueckelberg mass, evades the constraints from self-interacting DM with long-range force, while the gravitationally produced longitudinal dark photons, which decouple from the dark fermions, remain negligible both as dark radiation and as DM. We also verified that sequential production of dark fermions via dark-photon annihilation is negligible.} More generally, our results illustrate the fact that the gravitational portal
to an interacting hidden sector could be dominated by quantum anomaly effects, through channels which survive even lighter dark fermions. 

{We also considered the possibility that the hidden sector reaches thermal equilibrium at a temperature $T'\ll T$, in which case the gravitationally produced abundance no longer applies. Gravitational production then sets a lower bound on $T'/T$ for a given reheating temperature, to which the anomaly-induced production from the inflaton condensate contributes in addition to the production from the SM thermal bath. It could be of interest to study further the dark matter production in the case of  thermalization from the gravitational portal.}

\section*{Acknowledgements}
We thank Yann Mambrini for discussions. SC is supported by a Humboldt Research Fellowship of the Alexander von Humboldt Foundation. The work of MT is supported by the belgian FRS/FNRS, the BLU-ULB Brussels laboratory of the Universe and the IISN convention No. 4.4503.15. The work of AK was supported by a ULB postdoctoral grant. This work was made possible by Institut Pascal at Université Paris-Saclay with the support of the program “Investissements d’avenir” ANR-11-IDEX-0003-01, the P2I axis of the Graduate School of Physics of Université Paris-Saclay, as well as IJCLab, CEA and ACME.

\appendix

\section{Three-body phase space for Dalitz process}
\label{appendix_3body}
For the annihilation of two non-relativistic inflaton quanta, the total incoming four-momentum is therefore
\begin{equation}
    P^\mu = p_1^\mu+p_2^\mu
    =
    (2m_\phi,\mathbf{0}),
    \qquad
    s=P^2=4m_\phi^2 .
\end{equation}
The phase-space measure itself is purely kinematic and does not depend on the
interpretation of the initial state as a condensate. 
For a generic three-body final state,
\begin{equation}
    P \longrightarrow p_3+p_4+k ,
\end{equation}
the Lorentz-invariant three-body phase-space measure is
\begin{equation}
    d\Phi_3
    =
    \prod_{i}
    \frac{d^3\mathbf{p}_i}{(2\pi)^3\,2E_i}
    (2\pi)^4
    \delta^{(4)}
    \left(
        P-\sum_{i}p_i
    \right).
    \label{eq:PS3_general}
\end{equation}
Working in the center-of-mass frame,
\begin{equation}
    P^\mu=(\sqrt{s},\mathbf{0}),
\end{equation}
We first use the three-momentum delta function to set $\mathbf{p}_3$, and then choosing $\mathbf{p}_3$ as the polar axis, the integration over its solid angle gives $4\pi$, while the azimuthal angle of $\mathbf{p}_4$ gives an additional factor $2\pi$.
Equation~\eqref{eq:PS3_general} then becomes
\begin{equation}
    d\Phi_3
    =
    \frac{p_3^2 p_4^2\,dp_3\,dp_4}
         {32\pi^3 E_3E_4E_k}
    d\cos\theta_{34}\,
    \delta\left(
        \sqrt{s}-E_3-E_4-E_k
    \right).
    \label{eq:PS3_before_angle}
\end{equation}
where $\theta_{34}$ is the angle between $\mathbf{p}_3$ and $\mathbf{p}_4$.
The remaining angular integral can be performed using the energy-conservation
delta function. Consequently,
\begin{equation}
    \int_{-1}^{1}
    d\cos\theta_{34}\,
    \frac{1}{E_k}
    \delta\left(
        \sqrt{s}-E_3-E_4-E_k
    \right)
    =
    \frac{1}{p_3p_4}.
    \label{eq:PS3_delta}
\end{equation}
The three-body phase-space measure therefore takes the simple form
\begin{equation}
    d\Phi_3
    =
    \frac{dE_3\,dE_4}{32\pi^3}
    \label{eq:PS3_energy}
\end{equation}
with the masses of the final-state particles entering only through the kinematically allowed integration region.\\

For the anomaly-induced Dalitz process considered in
section~\ref{subsec:anomaly_dalitz},

\begin{equation}
    \phi\phi
    \to
    \gamma'(k)\,
    \chi(p_3)\,
    \bar{\chi}(p_4),
\end{equation}
It is convenient to introduce the invariant mass of the fermion-antifermion pair,

\begin{equation}
    q\equiv p_3+p_4,
    \qquad
    z\equiv\frac{q^2}{s},
    \qquad
    r\equiv\frac{m_\chi}{m_\phi}.
\end{equation}
Since the remaining dark photon is considered massless,
\begin{equation}
    E_k
    =
    \frac{s-q^2}{2\sqrt{s}}
    =
    m_\phi(1-z),
\end{equation}
and energy conservation gives
\begin{equation}
    E_3+E_4
    =
    m_\phi(1+z).
\end{equation}
At fixed $z$, the fermion energy is bounded by
\begin{equation}
    E_3^\pm
    =
    \frac{m_\phi}{2}
    \left[
        1+z
        \pm
        (1-z)
        \sqrt{1-\frac{r^2}{z}}
    \right],
    \label{eq:E3_boundaries}
\end{equation}
with
\begin{equation}
    r^2\leq z\leq1.
\end{equation}
The three-body phase space measure can thus equivalently be written as
\begin{equation}
    d\Phi_3
    =
    \frac{m_\phi}{32\pi^3}\,
    dE_3\,dz
    \label{eq:PS3_Ez}
\end{equation}
For the Dalitz process considered in section~\ref{subsec:anomaly_dalitz}, we defined
\begin{equation}
    I_\chi^{\rm D}
    \equiv
    N_\chi\int d\Phi_3
    \sum_{\rm spin,\,pol.}
    |\mathcal{M}_{\rm D}^{0\frac{1}{2}}|^2 .
\end{equation}
Using Eq.~\eqref{eq:PS3_Ez}, its differential form is therefore
\begin{equation}
    \frac{d I_\chi^{\rm D}}{dz}
    = N_\chi
    \frac{m_\phi}{32\pi^3}
    \int_{E_3^-}^{E_3^+}
    dE_3\,
    \sum_{\rm spin,\,pol.}
    |\mathcal{M}_{\rm D}^{0\frac{1}{2}}|^2
    \label{eq:dID_dz_appendix}
\end{equation}

For completeness, for a genuine one-particle decay of mass $M$ one has
\begin{equation}
    \Gamma_{1\rightarrow3}
    =
    \frac{1}{2M}
    \int d\Phi_3\,
    |\mathcal{M}|^2,
\end{equation}
whereas for the annihilation of two non-relativistic particles of mass
$m_\phi$,
\begin{equation}
    \sigma v_{\rm rel}
    =
    \frac{1}{4m_\phi^2}
    \int d\Phi_3\,
    |\mathcal{M}|^2.
    \label{eq:sigmav_2to3}
\end{equation}
For the coherent inflaton condensate,
the corresponding particle-production rate is instead written as
\begin{equation}
    R
    =
    \mathcal N\frac{\rho_\phi^2}{16m_\phi^4}
    \int d\Phi_3\,|\mathcal{M}|^2 ,
\end{equation}
with $\mathcal N$ the number of particles of one species produced in the final state. Using Eq.~\eqref{eq:sigmav_2to3} and $n_\phi=\rho_\phi/m_\phi$, this may equivalently be written as
\begin{equation}
    R
    =
    \mathcal N\frac{1}{4}\,
    n_\phi^2\,
    \sigma v_{\rm rel}
\end{equation}
The factor differs from the usual dilute-gas expression, as discussed in section~\ref{subsec:production_rates}.\\

\section{Sequential production of dark fermions}
\label{appendix:secondary}
The anomaly-induced gravitational portal produces a population of dark photons that may subsequently annihilate into fermion-antifermion pairs \cite{Hambye:2019dwd}
\begin{equation}
    \gamma'\gamma' \rightarrow \chi\bar{\chi}
    \label{eq:secondary_process}
\end{equation}
and could provide an additional source of DM. We verify in this appendix that this contribution is sub-leading. 

The dark photons number density at the end of reheating follows Eq.(\ref{eq:ngamma_RH}), and their associated energy density Eq.(\ref{eq:rho_gamma_RH}). From this, we can estimate the sequential production of dark fermions induced by
Eq.~\eqref{eq:secondary_process}. The corresponding source term can be written as
\begin{equation}
    R_\chi^{\rm seq}
    =
    n_{\gamma'_T}^2
    \left\langle
        \sigma v
    \right\rangle_{\gamma'\gamma'\rightarrow\chi\bar{\chi}} 
    \label{eq:R_secondary}
\end{equation}
For relativistic dark photons with $E_{\gamma'}\gg m_\chi$, the annihilation cross section can be
estimated parametrically as \cite{Breit:1934zz},
\begin{eqnarray}
\sigma_{\gamma'\gamma'\to\chi\bar\chi}&\sim&N_\chi\frac{4\pi\alpha'^2}{s}\left[\log\frac{s}{m_\chi^2}-1\right],
  \label{eq:sigma_gamma_gamma}
\end{eqnarray}
Averaging over an isotropic population of dark photons of energy $E_{\gamma'}$ and using Eq.~\eqref{eq:Rdalitz_smallr} and
Eq.~\eqref{eq:ngamma_RH} to compare with the Dalitz source gives
\begin{equation}
  \frac{R_\chi^{\rm seq}}{R_\chi^{\rm D}}\bigg|_{a_{\rm end}}\sim\frac{N_\chi^2\alpha'^3}{144\pi}\,
  \frac{\rho_{\rm end}}{M_P^2\,m_\phi^2} \simeq1.5\times10^{-7}\,N_\chi^2e'^6
  \ll1 
  \label{eq:Rsec_over_RD}
\end{equation}
for $m_\phi = 3\times10^{13}\,\rm GeV$ and $\rho_{\rm end}=(5.2\times 10^{15}\, \rm GeV)^4$. Since $n_{\gamma'_T}\propto a^{-3}$ while $\sigma\propto E_{\gamma'}^{-2}\propto a^2$, the comoving yield keeps growing as $a$ after
reheating, until $E_{\gamma'}\sim m_\chi$. Integrating up to this time gives
\begin{equation}
  \frac{Y_\chi^{\rm seq}}{Y_\chi^{\rm D}}\sim0.08\,N_\chi^2\,e'^6
  \left(\frac{10^{7}\,{\rm GeV}}{m_\chi}\right)
  \left(\frac{T_{\rm RH}}{10^{14}\,{\rm GeV}}\right)^{2/3}
\end{equation}
which reaches at most $\mathcal O(10\%)$ for $e'=1$ (and $N_\chi = 1$) and $m_\chi\sim10^{7}$~GeV, and is negligible elsewhere.

In addition to the source coming from the inflaton condensate, dark photons are also gravitationally produced by scatterings in the visible thermal bath, ${\rm SM}+{\rm SM}\to h^*\to\gamma'\gamma'$, which are not subject to mass suppression \cite{Bernal:2018qlk,Mambrini:2021zpp,Clery:2021bwz}. However, their abundance $n_{\gamma'}^{\rm bath}$, is generally smaller than the one of dark photons produced by the anomaly (see Eq.(\ref{eq:ngamma_bath})), and their subsequent conversion into fermions through $\gamma'\gamma'\to\chi\bar\chi$ involves an additional scattering in the dilute dark sector population. This is parametrically suppressed with respect to the direct gravitational production of fermions from the thermal bath, already considered before. We therefore can neglect it.

If dark photons are already massive during reheating, an additional longitudinal population is gravitationally produced without mass suppression, as discussed in section \ref{subsec:DP_mass}. Although the population of gravitationally produced longitudinal dark photons can be large, in the parameter region where its contribution to DM relic remains negligible, see Eq.(\ref{eq:mLreh}), the sequential production also remains negligible. Indeed, for the conserved dark QED current, annihilation channels to dark electrons involving one or two longitudinal dark photons $\gamma'_L$ are suppressed by powers of $m_{\gamma'}/E_{\gamma'}$.  

The sequential contribution remains thus subleading in the regime considered. The gravitationally produced dark sector is highly underoccupied, $n_{\gamma'_T}/E_{\gamma'}^3\ll1$, so that dark-photon reannihilation is a slow process. Consequently, in the regime where the dark sector does not interact sizably after being produced, sequential fermion production can consistently be neglected. 

\section{Feynman rules}
\label{app:FR}
\enlargethispage{2\baselineskip}
We give for reference the relevant Feynman rules \cite{Han:1998sg}. The contribution of the trace anomaly is discussed in the body of the text. 

The de Donder harmonic-gauge propagator for the massless graviton is given by
\begingroup
\interdisplaylinepenalty=10000
\bea
&&\begin{tikzpicture}[baseline={-0.1cm}]
  \begin{feynman}[every blob={/tikz/fill=gray!30,/tikz/inner sep=2pt}]
    \vertex (L) at (0,0) {$\mu\nu$};
    \vertex (R) at (4,0) {$\alpha\beta$};
    \vertex (C1) at (1,0) ;
    \vertex (C2) at (3,0) ;
    \diagram*{
      (L) -- [draw=none, momentum'={[arrow shorten=0.4]}, edge label=$P$] (R),
      (C1) -- [graviton, style=black] (C2)
    };
  \end{feynman}
\end{tikzpicture} \nonumber\\
&=&  
\displaystyle \frac{i}{P^2} \frac{1}{2} \left( \eta^{\mu \alpha} \eta^{\nu \beta} + \eta^{\mu \beta} \eta^{\nu \alpha} - \eta^{\mu \nu} \eta^{\alpha \beta} \right)
\eea
while the vertices read,
\enlargethispage{2\baselineskip}
\be
\begin{tikzpicture}[baseline={-0.1cm}]
  \begin{feynman}[every blob={/tikz/fill=gray!30,/tikz/inner sep=2pt}]
    \vertex (c1) at (1., 0.) {\(\mu\)} ;
    \vertex (c2) at (2., 0.) ;
    \vertex (f1) at (3.,0.75) ;
    \vertex (f2) at (3.,-0.75) ;
    \diagram* {
   (c1) -- [draw=none, text=black, momentum'={[arrow shorten=0.2]}, edge label=\(k\)](c2),
      (c1) -- [boson, style=black] (c2),
      (f1) -- [anti fermion,momentum'={[arrow shorten=0.3]\(p_2\)}] (c2),
      (f2) -- [fermion,momentum={[arrow shorten=0.3]\(p_1\)}] (c2),      
    };
  \end{feynman}
\end{tikzpicture}
= {i e'\gamma^\mu} 
    \ee

\bea
&&\begin{tikzpicture}[baseline={-0.1cm}]
  \begin{feynman}[every blob={/tikz/fill=gray!30,/tikz/inner sep=2pt}]
    \vertex (c1) at (1., 0.) {\(\mu\nu\)} ;
    \vertex (c2) at (2., 0.) ;
    \vertex (s1) at (3.,0.75) ;
    \vertex (s2) at (3.,-0.75) ;
    \diagram* {
   (c1) -- [draw=none, text=black, momentum'={[arrow shorten=0.2]}, edge label=\(P\)](c2),
      (c1) -- [graviton, style=black] (c2),
      (s1) -- [scalar,momentum'={[arrow shorten=0.3]\(p_2\)}] (c2),
      (s2) -- [scalar,momentum={[arrow shorten=0.3]\(p_1\)}] (c2),      
    };
  \end{feynman}
\end{tikzpicture}
= {i \over M_P}\left[ p_1^\mu p_2^\nu + p_1^\nu p_2^\mu\right.
 \\
&& \qquad\qquad\qquad\qquad - \left.\eta^{\mu\nu} \left(p_1\cdot p_2 + m_\phi^2\right)\right]\nonumber
\eea
  \bea
&&\begin{tikzpicture}[baseline={-0.1cm}]
  \begin{feynman}[every blob={/tikz/fill=gray!30,/tikz/inner sep=2pt}]
    \vertex (c1) at (1., 0.) {\(\mu\nu\)} ;
    \vertex (c2) at (2., 0.) ;
    \vertex (f1) at (3.,0.75) ;
    \vertex (f2) at (3.,-0.75) ;
    \diagram* {
   (c1) -- [draw=none, text=black, momentum'={[arrow shorten=0.2]}, edge label=\(P\)](c2),
      (c1) -- [graviton, style=black] (c2),
      (f1) -- [anti fermion,momentum'={[arrow shorten=0.3]\(p_2\)}] (c2),
      (f2) -- [fermion,momentum={[arrow shorten=0.3]\(p_1\)}] (c2),      
    };
  \end{feynman}
\end{tikzpicture}
= -{i \over M_P}\left[ {1\over 4} \gamma^\mu (p_1- p_2)^\nu \right. \\
&&\qquad + {1\over 4} \gamma^\nu (p_1- p_2)^\mu 
- \left.\eta^{\mu\nu} \left({1\over 2} (\fsl p_1 - \fsl p_2) - m_\chi\right)\right]\nonumber
\eea
\newpage
\bea
&&\begin{tikzpicture}[baseline={-0.1cm}]
  \begin{feynman}[every blob={/tikz/fill=gray!30,/tikz/inner sep=2pt}]
    \vertex (c1) at (1., 0.) {\(\mu\nu\)} ;
    \vertex (c2) at (2., 0.) ;
    \vertex (f1) at (3.,0.75) ;
    \vertex (f2) at (3.,-0.75) ;
    \vertex (a1) at (3.25,0.) {\(\alpha\)};
    \diagram* {
   (c1) -- [draw=none, text=black, momentum'={[arrow shorten=0.2]}, edge label=\(P\)](c2),
      (c1) -- [graviton, style=black] (c2),
      (f1) -- [anti fermion,momentum'={[arrow shorten=0.3]\(p_2\)}] (c2),
      (f2) -- [fermion,momentum={[arrow shorten=0.3]\(p_1\)}] (c2), 
      (a1) -- [boson] (c2),
    };
  \end{feynman}
\end{tikzpicture} \quad \mbox{\rm with}\quad D_\mu = \partial_\mu  - i e^\prime A_\mu
\\
&& \qquad = {i e^\prime \over M_P} \left[\eta^{\mu \nu}  \gamma^\alpha- {1\over 2} \left( \gamma^\mu \eta^{\nu \alpha}+
\gamma^\nu \eta^{\mu \alpha}\right)   \right]\nonumber
\eea
\bea
&&
\begin{tikzpicture}[baseline={-0.1cm}]
  \begin{feynman}[every blob={/tikz/fill=gray!30,/tikz/inner sep=2pt}]
    \vertex (c1) at (1., 0.) {\(\mu\nu\)} ;
    \vertex (c2) at (2., 0.) ;
    \vertex (a1) at (3.,0.75) {\(\beta\)};
    \vertex (a2) at (3.,-0.75) {\(\alpha\)};
    \diagram* {
   (c1) -- [draw=none, text=black, momentum'={[arrow shorten=0.2]}, edge label=\(P\)](c2),
      (c1) -- [graviton, style=black] (c2),
      (a1) -- [boson,momentum'={[arrow shorten=0.3]\(k_2\)}] (c2),
      (a2) -- [boson,momentum={[arrow shorten=0.3]\(k_1\)}] (c2),      
    };
  \end{feynman}
\end{tikzpicture}
= {i \over M_P}\left[ - (k_1^\mu k_2^\nu + k_1^\nu k_2^\mu)\eta^{\alpha \beta}\right.\\
&& \quad \qquad +  \left. (\eta^{\nu \beta} k_1^\mu  + \eta^{\mu \beta} k_1^\nu ) k_2^\alpha +  (\eta^{\nu \alpha} k_2^\mu  + \eta^{\mu \alpha} k_2^\nu ) k_1^\beta \right.\nonumber\\
&& \quad \qquad - (k_1\cdot k_2 + m_{\gamma'}^2)(\eta^{\mu \alpha}\eta^{\nu \beta}+ \eta^{\mu \beta}\eta^{\nu \alpha})\nonumber\\
&& \quad \qquad + \left. \eta^{\mu \nu}\left((k_1\cdot k_2 +m_{\gamma'}^2)\eta^{\alpha \beta}-k_1^\beta k_2^\alpha\right)\right]\nonumber
\eea
\endgroup

\setlength{\bibsep}{0pt}
\bibliographystyle{JHEP}
\bibliography{biblio}

\end{document}